\documentclass[twocolumn,nofootinbib,superscriptaddress]{revtex4-2}
\usepackage[T1]{fontenc} 
\usepackage{hyperref}
\usepackage{bm}
\usepackage{graphicx}
\usepackage{braket}
\usepackage{aas_macros}
\usepackage{amsmath}
\usepackage{amssymb}
\usepackage{color}

\newcommand{\BG}{{\scriptscriptstyle\rm BG}}

\begin{document}

\hfill {\scriptsize RBI-ThPhys-2025-36, YITP-25-147}

\title{Post-collapse Lagrangian perturbation theory in three dimensions}
\author{Shohei Saga}
\email{shohei.saga@yukawa.kyoto-u.ac.jp}
\affiliation{Institute for Advanced Research, Nagoya University, Furo-cho Chikusa-ku, Nagoya 464-8601, Japan}
\affiliation{Kobayashi-Maskawa Institute for the Origin of Particles and the
Universe, Nagoya University, Chikusa-ku, Nagoya, 464-8602, Japan}
\affiliation{Sorbonne Universit\'e, CNRS, UMR7095, Institut d'Astrophysique de Paris, 98bis boulevard Arago, F-75014 Paris, France}
\author{St{\'e}phane Colombi}
\affiliation{Sorbonne Universit\'e, CNRS, UMR7095, Institut d'Astrophysique de Paris, 98bis boulevard Arago, F-75014 Paris, France}
\author{Atsushi Taruya}
\affiliation{Center for Gravitational Physics and Quantum Information, Yukawa Institute for Theoretical Physics, Kyoto University, Kyoto 606-8502, Japan}
\affiliation{Kavli Institute for the Physics and Mathematics of the Universe (WPI), Todai institute for Advanced Study, University of Tokyo, Kashiwa, Chiba 277-8568, Japan}
\affiliation{Korea Institute for Advanced Study, 85 Hoegiro, Dongdaemun-gu, Seoul 02455, Republic of Korea}
\author{Cornelius Rampf}
\affiliation{Division of Theoretical Physics, Ru\dj er Bo\v{s}kovi\'c Institute, Bijeni\v{c}ka cesta 54, 10000 Zagreb, Croatia}
\affiliation{Department of Astrophysics, University of Vienna, T\"{u}rkenschanzstra{\ss}e 17, 1180 Vienna, Austria}
\affiliation{Faculty of Mathematics, University of Vienna, Oskar-Morgenstern-Platz 1, 1090 Vienna, Austria}

\author{Abineet Parichha}
\affiliation{Sorbonne Universit\'e, CNRS, UMR7095, Institut d'Astrophysique de Paris, 98bis boulevard Arago, F-75014 Paris, France}

\date{\today}

\begin{abstract}
The gravitational collapse of collisionless matter leads to shell-crossing singularities that challenge the applicability of standard perturbation theory. Here, we present the first fully perturbative approach in three dimensions by using Lagrangian coordinates that asymptotically captures the highly nonlinear nature of matter evolution after the first shell-crossing. This is made possible essentially thanks to two basic ingredients: (1) We employ high-order standard Lagrangian perturbation theory to evolve the system until shell-crossing, and (2) we exploit the fact that the density caustic structure near the first shell-crossing begins generically with pancake formation. The latter property allows us to exploit largely known one-dimensional results to determine perturbatively the gravitational backreaction after collapse, yielding accurate solutions within our post-collapse perturbation theory (PCPT) formalism. We validate the PCPT predictions against high-resolution Vlasov-Poisson simulations and demonstrate that PCPT provides a robust framework for describing the early stages of post-collapse dynamics.

\end{abstract}

\maketitle

\section{Introduction}

Cold dark matter (CDM), the dominant matter component in the standard cosmological model, drives the formation and evolution of large-scale structures in the Universe~\cite{1982ApJ...263L...1P,1984ApJ...277..470P,1984Natur.311..517B}. As a collisionless and pressureless component, and assuming the validity of the continuum limit, CDM obeys the Vlasov-Poisson equations, which govern its dynamics in six-dimensional phase space~\cite{1980lssu.book.....P}. Due to its cold nature, the phase-space distribution of CDM can be approximated as a three-dimensional hypersurface evolving in six-dimensional phase space. In this framework, the matter fluid is initially single stream which comes with zero velocity dispersion. Gravitational \mbox{(self-)}interactions, however, can lead to a significant accumulation of matter at focused locations, implying that matter trajectories cross, leading to a local proliferation of multiple fluid streams that generate non-zero velocity dispersion. The instant of bifurcation between single- and multistreaming in phase space is called shell-crossing.

In the weakly nonlinear regime (before shell-crossing), perturbative techniques can  describe accurately the matter evolution. Within this framework, 
the two base approaches are standard perturbation theory (SPT)~\citep[see e.g.,][for a review]{2002PhR...367....1B} and Lagrangian perturbation theory (LPT)~\citep[see e.g.,][]{1970A&A.....5...84Z,1989RvMP...61..185S,1992MNRAS.254..729B,1995A&A...296..575B,Rampf:2012xa}.
Both SPT and LPT have been widely applied to model statistical observables in large-scale structure \citep[see, e.g.,][]{2006PhRvD..73f3520C,2013PhRvD..87h3522V,2008PhRvD..77b3533C,2008ApJ...674..617T,2008PhRvD..77f3530M,2008JCAP...10..036P,2009PhRvD..80l3503T,2011PhRvD..83h3518M,2012PhRvD..85l3519B,2013PhRvD..87h3522V}.  
In SPT, the Eulerian frame of reference is used and the density contrast acts as a small perturbative parameter, assuming a single-stream flow. In contrast, LPT tracks the trajectories of matter elements from their initial Lagrangian positions and expands the displacement field perturbatively. Astonishingly, LPT is able to predict the emergence of shell-crossing and, to some extent, multi-streaming, which SPT cannot. Nevertheless, both approaches face a fundamental limitation: their perturbative expansions substantially get worse once shell crossing occurs, leading to a rapid loss of predictive power in the multistream regime~(see e.g., Refs.~\cite{2014PhRvD..89b3502B,2014JCAP...01..010B,2016PhLB..762..247N,2020MNRAS.499.1769H} for SPT and Refs.~\cite{2014JFM...749..404Z,2015MNRAS.452.1421R,Rampf:2021rqu,2021MNRAS.500..663M,2022A&A...664A...3S,2023PhRvD.107b3515R} for LPT).

Understanding the nonlinear dynamics of structure formation, particularly in multistream regions, remains a major challenge in cosmology.
To address this challenge, several theoretical frameworks have been developed in recent years. Among them, the effective field theory (EFT) of the large-scale structure has provided a systematic procedure to incorporate small-scale nonlinearities into large-scale fluid equations, albeit with free parameters that typically require calibration against simulations~\cite{2012JCAP...07..051B,2012JHEP...09..082C,2014PhRvD..89d3521H,2015PhRvD..92l3007B,2020JCAP...05..005D}. More recently, Vlasov perturbation theory (VPT) has been developed as a first-principles approach that directly tackles the Vlasov-Poisson system, employing the cumulant expansion of the phase-space distribution function~\cite{2023PhRvD.107f3539G,2023PhRvD.107f3540G}. VPT offers a rigorous method to describe CDM dynamics incorporating higher-order moments of the distribution function and has shown promising results~\cite{2025arXiv250220451G}, with a few parameters related to the ``background'' dispersions. Both EFT and VPT aim to provide controlled extensions of SPT into the nonlinear regime by treating the multi-stream responses at the statistical level; by contrast, in this paper we aim for a description that is also meaningful at the deterministic level.

An alternative and highly insightful Lagrangian approach is provided by the post-collapse perturbation theory (PCPT) formalism developed in Refs.~\cite{2015MNRAS.446.2902C,2017MNRAS.470.4858T,2021MNRAS.505L..90R} (see also \cite{2018JCAP...06..028P}) for one-dimensional dynamics, i.e., in the case of infinite massive sheets evolving along one direction in the expanding three-dimension universe. This method analytically describes post-shell-crossing dynamics by computing the gravitational backreaction from multistream flows onto first-order LPT, known as the Zel'dovich approximation~\cite{1970A&A.....5...84Z}, which is an exact solution for the pre-collapse regime in 1D. In the 1D case, PCPT in its current form can already predict the second shell-crossing, and is able to reproduce qualitatively the phase-space structures and density profiles measured in simulations even slightly beyond the second shell crossing.  The strength of PCPT lies in its ability to handle the nonlinearities induced by shell crossing while remaining entirely analytical, thereby bridging the gap between Zel'dovich flow and nonlinear multi-stream flow at a tractable level.

Extending PCPT from one- to three-dimensional dynamics, however, presents significant technical and conceptual challenges. For instance, the geometry of multistream regions becomes considerably more complex, and shell crossing occurs generically along different axes at different times. Despite these challenges, a perturbative treatment of post-collapse dynamics in three dimensions is crucial for improving our understanding of cosmic structure formation. Motivated by these considerations, we address this problem by focusing on the dynamics of symmetrical pancake collapse. We develop a fully analytical three-dimensional extension of post-collapse perturbation theory building upon our previous investigations in one dimension~\cite{2018PhRvL.121x1302S,2022A&A...664A...3S,2023A&A...678A.168S}. Our approach generalizes one-dimensional PCPT by perturbatively computing the gravitational backreaction from multistream regions around a pre-collapse `background flow', where the latter is modelled using fixed-order LPT. By exploiting the specific pancake structure, in which collapse occurs predominantly along one direction, we reduce the problem to an effectively one-dimensional treatment along the collapse axis while retaining the transverse directions as parameters. This allows us to derive explicit corrections to the displacement and velocity fields beyond shell crossing. As a benchmark, we test our predictions against Vlasov-Poisson simulations performed with the public code \texttt{ColDICE}~\cite{2016JCoPh.321..644S,2021A&A...647A..66C}, demonstrating improved performance of PCPT over standard LPT in the multistream regime.

This paper is structured as follows. Section~\ref{sec: basic eq} introduces the basic equations of motion in Lagrangian coordinates. In Sec.~\ref{sec: pcpt}, we develop the analytical methods for describing the gravitational dynamics within multistream regions. We derive expressions for the gravitational backreaction. The PCPT predictions are examined through comparisons with high-resolution Vlasov-Poisson \texttt{ColDICE} simulations in Sec.~\ref{sec: results}. Section~\ref{sec: summary} is devoted to summarizing the key results of this work and discussing future directions. Finally, appendices \ref{sec: app Fx}, \ref{sec: app derivations}, and \ref{sec: app LPT dependence} provide the details of the derivations of the gravitational force from the multistream region, the derivations of the corrections to the background flow, and discuss the dependence of the results on the choice of the background flow (high-order LPT solutions), respectively.

\section{Basic equations\label{sec: basic eq}}

We consider the Lagrangian equations of motion of a matter element \citep[e.g.,][]{1980lssu.book.....P}:
\begin{align}
& \frac{{\rm d}\bm{x}}{{\rm d}t}  = \frac{\bm{v}}{a} , \label{eq:vel}\\
& \frac{{\rm d} \bm{v}}{{\rm d}t} + H \bm{v} = -\frac{1}{a}\bm{\nabla}_{x} \phi(\bm{x}), 
\label{eq:basic_EoM}\\
& \nabla^{2}_{x}\phi(\bm{x}) = 4\pi G a^{2}\,\bar{\rho} \delta(\bm{x}) . 
\label{eq:Poisson}
\end{align}
where the quantities $\bm{x}$, $\bm{v}$, $a$, $H(t)=a^{-1} {\rm d}a/{\rm d}t$, $\bm{\nabla}_{x}$, $\phi$, $\bar{\rho}$, and $\delta = \rho/\bar{\rho}-1$ are the Eulerian comoving position, the peculiar velocity, the scale factor of the Universe, the Hubble parameter, the spatial gradient operator in Eulerian space, the Newton gravitational potential, the background mass density, and the matter density contrast, respectively. For convenience, we rewrite the set of equations in terms of the super-conformal time~\cite{1973Ap......9..144D,1998MNRAS.297..467M}: ${\rm d}\tau = a^{-2}{\rm d}t$. Then, Eqs.~(\ref{eq:vel})--(\ref{eq:Poisson}) become
\begin{align}
\frac{{\rm d}\bm{x}}{{\rm d}\tau} &= \bm{u} , \label{eq:vel2}\\
\frac{{\rm d}\bm{u}}{{\rm d}\tau} &=-\bm{\nabla}_{x}\Phi , \label{eq:basic_EoM2}\\
\nabla^{2}_{x}\Phi &= \frac{3}{2}\Omega_{\rm m0}H^{2}_{0}a\delta(\bm{x}) , \label{eq:Poisson2}
\end{align}
where $\Omega_{\rm m0}$ and $H_0$ are the present-day values of the matter density and Hubble parameter, respectively, and
we have defined $\Phi = a^{2} \phi$ and $\bm{u} = a\bm{v}$.

We consider the Lagrangian description, which relates the Lagrangian position (initial position), $\bm{q}$, of each mass element to the Eulerian position at a given time, $\bm{x}(\tau,\bm{q})$, through the displacement field, $\bm{\psi}(\tau,\bm{q})$:
\begin{align}
\bm{x}(\tau, \bm{q}) = \bm{q} + \bm{\psi}(\tau,\bm{q}) . \label{eq:x-Psi}
\end{align}
With this description, the velocity of each mass element is given by 
\begin{align}
\bm{u} = \frac{{\rm d}\bm{\psi}}{{\rm d}\tau} . \label{eq:u-Psi}
\end{align}
Using Eqs.~(\ref{eq:vel2}) and (\ref{eq:basic_EoM2}) in the Lagrangian description, the solution of the equations of motion is formally given by
\begin{align}
\bm{x}(\tau,\bm{q}) & = \bm{x}(\tau_{\rm ini},\bm{q}) + \delta \bm{x}(\tau, \bm{q}), \label{eq: x formal} \\
\bm{u}(\tau,\bm{q}) & = \bm{u}(\tau_{\rm ini},\bm{q}) + \delta \bm{u}(\tau,\bm{q}), \label{eq: u formal}
\end{align}
where the corrections $\delta \bm{x}$ and $\delta \bm{u}$ are given by
\begin{align}
\delta \bm{x}(\tau,\bm{q}) & = \int^{\tau}_{\tau_{\rm ini}}{\rm d}\tau'\, \bm{u}(\tau',\bm{q}), \label{eq: Delta x formal}\\
\delta \bm{u}(\tau,\bm{q}) & = - \int^{\tau}_{\tau_{\rm ini}}{\rm d}\tau'\, \bm{\nabla}_{x}\Phi(\tau', \bm{x}(\tau',\bm{q})). \label{eq: Delta u formal}
\end{align}
The initial time $\tau_{\rm ini}$ will be specified in the next section.

Recalling the mass conservation between the Lagrangian space and Eulerian space, $(1+\delta(\bm{x})){\rm d}^{3}\bm{x} = {\rm d}^{3}\bm{q}$, we have
\begin{align}
1 + \delta(\bm{x}) = \frac{1}{J}  \label{eq:jac}
\end{align}
before shell crossing and assuming initial homogeneity, with $J = \det\left(J_{ij}\right)$ being the Jacobian of the matrix $J_{ij} = \left[ \partial \bm{x}/\partial \bm{q}\right]_{ij}$. We define the shell crossing time, $\tau_{\rm sc}$, by the moment when the Jacobian $J$ first becomes zero, i.e., $J(\tau_{\rm sc},\bm{q}_{\rm sc}) = 0$.

We note that, until the first shell-crossing time, assuming that the displacement field is small, we can employ LPT to perturbatively solve the set of equations (\ref{eq:vel2}), (\ref{eq:u-Psi}) and (\ref{eq:Poisson2}) \cite{1989RvMP...61..185S,1992ApJ...394L...5B,1992MNRAS.254..729B,1993MNRAS.264..375B,1995A&A...296..575B,1994ApJ...427...51B}.  In the next section, we develop the PCPT formalism to describe the fluid motion in three-dimensional space after the first shell crossing, but without yet assuming any pre-collapse modelling such as LPT. We will return to LPT again in Sec.~\ref{sec: results} as a specific application of PCPT.

\section{Post-collapse perturbation theory\label{sec: pcpt}}

In this section, we develop the PCPT formalism by extending previous works of Refs.~\cite{2015MNRAS.446.2902C,2017MNRAS.470.4858T} to three dimensions. 
The procedure for formulating the PCPT is outlined as follows:
\begin{enumerate}
    \item \textit{System setup:} We focus on a symmetric proto-pancake seeded by a locally axisymmetric motion. This configuration leads to the first shell-crossing occurring along a single, well-defined axis.
    \item \textit{Background flow:} To describe the post-collapse dynamics, we introduce a {\em background flow}, which describes the post-collapse motion without the backreaction from the multi-stream region. Assuming that the multistream region is small, the background flow is Taylor expanded in Lagrangian space around the position of the first shell-crossing point. By expanding the background flow to linear order in time with respect to collapse time, we explicitly derive the expressions for the key quantities needed to perturbatively integrate corrections to the equations of motion.
    \item \textit{Gravitational backreaction:} Building on the calculations of Ref.~\cite{2023A&A...678A.168S}, we fully account for the gravitational backreaction to derive an (asymptotic) analytical expression for the force field expected shortly after shell-crossing in the multi-stream region and its close vicinity.
    \item \textit{Positions and velocities in PCPT:} The obtained gravitational backreaction enables the perturbative calculation of corrections to the equations of motion, leading to the expressions for the positions and velocities in the PCPT framework.
\end{enumerate}
The remainder of this section is organized as follows. Sec.~\ref{sec: setup} introduces the symmetric proto-pancake system and the concept of the {\em background flow}.
In Sec.~\ref{sec: caustic}, we investigate the caustic structure of the multistream region shortly after shell crossing.
In Sec.~\ref{sec: force}, we present the expression for the force field that accounts for the backreaction from the multi-stream region, based on the work of Ref.~\cite{2023A&A...678A.168S}. 
In Sec.~\ref{sec: x exp}, we derive expressions of the key quantities needed to integrate perturbatively corrections to the equations of motion.
Finally, Sec.~\ref{sec: formulation}, presents the main result of this paper, namely the final expressions for the positions and velocities in the PCPT framework.

\subsection{
Setup and background flow
 \label{sec: setup}}

In one-dimensional gravitational dynamics, the Zel'dovich solution is exact until shell-crossing time. By calculating in a perturbative way the backreaction from the multi-stream region of the extrapolated Zel'dovich flow, here dubbed \textit{the background flow}, and by adding the corrections to this background flow, Ref.~\cite{2017MNRAS.470.4858T} constructed a theory describing very well the post-collapse motion for one-dimensional dynamics up to next crossing time. Building upon the perturbative framework developed in 1D, our goal here is to extend the treatment to the 3D case. Rather than specifying a particular form for the background flow, we formally consider a symmetric but still quite generic form of the Eulerian position $\bm{x}(\tau, \bm{q})$, which should describe the fluid motion accurately until shell crossing, and (at least) approximately after it. Later on, we will employ LPT solutions for the background flow.

Assuming that the collapsing region is small allows us to perform a Taylor expansion of the background flow in Lagrangian space around the shell-crossing point $\bm{q} \equiv \bm{q}_{\rm sc}$. Neglecting $O(q^{4})$ and higher order terms, we have
\begin{align}
{\bm{x}}^{\BG}(\tau,\bm{q}) &
\simeq 
\bm{x}^\BG(\tau,\bm{q}_{\rm sc})
+ \frac{\partial\bm{x}^\BG(\tau,\bm{q}_{\rm sc})}{\partial q_{i}} (q_{i}-q_{{\rm sc}, i})
\notag \\
&
+ \frac{1}{2} 
\frac{\partial^{2}\bm{x}^\BG(\tau,\bm{q}_{\rm sc})}{\partial q_{i}\partial q_{j}}
 (q_{i}-q_{{\rm sc}, i})(q_{j}-q_{{\rm sc}, j})
\notag \\
& 
+ \frac{1}{6}
\frac{\partial^{3}\bm{x}^\BG(\tau,\bm{q}_{\rm sc})}{\partial q_{i}\partial q_{j}\partial q_{k}}
\notag \\
& 
\qquad \times 
(q_{i}-q_{{\rm sc}, i})(q_{j}-q_{{\rm sc}, j})(q_{k}-q_{{\rm sc}, k})
, \label{eq: expand x}
\end{align}
where we adopt the Einstein summation convention for the subscripts $i,j,k$, which correspond to $x$, $y$, or $z$ coordinates.

Throughout this paper, we focus on a subclass of axisymmetric proto-pancakes contracting along all axes of motion, that is seeds of dark matter halos.  However, the calculations of this section are still valid for arbitrary axisymmetric pancakes, that is seeds of axisymmetric filaments (expanding along one axis, instead of contracting) or axisymmetric sheets (expanding along 2 axes), and generalization to non symmetric configurations is cumbersome but straightforward. Understanding the multistream dynamics of these structures is one of the essential steps for describing large-scale structure formation. This setup therefore provides us with insights into a more generic picture.

Hereinafter, without loss of generality, we set the shell-crossing point for pancake collapse to
\begin{align}
  \bm{q}_{\rm sc} = \bm{0}.
\end{align}
With this new origin for the coordinate system, the axisymmetric assumption simply reads
\begin{align}
  x^\BG(\tau,q_x,q_y,q_z) &=x^\BG(\tau,q_x,-q_y,q_z) \notag \\
                                              & =x^\BG(\tau,q_x,q_y,-q_z) \notag \\
                                              &=-x^\BG(\tau,-q_x,q_y,q_z), \\
  y^\BG(\tau,q_x,q_y,q_z) &=y^\BG(\tau,-q_x,q_y,q_z) \notag \\
                                              &=y^\BG(\tau,q_x,q_y,-q_z) \notag \\
                                              &=-y^\BG(\tau,q_x,-q_y,q_z),\\
  z^\BG(\tau,q_x,q_y,q_z) &=z^\BG(\tau,-q_x,q_y,q_z) \notag \\
                                              & =z^\BG(\tau,q_x,-q_y,q_z) \notag \\
                                              &=-z^\BG(\tau,q_x,q_y,-q_z).
\end{align}
Owing to the locally quasi one-dimensional nature of the collapse, the extent of the multi-stream region along the collapse direction (here the $x$ axis) is asymptotically infinitely smaller compared to that of the other axes just after shell crossing~\cite{2022A&A...664A...3S,2023A&A...678A.168S}, and equation (\ref{eq: expand x}) reduces to, in the symmetric configuration,
\begin{align}
x^\BG(\tau,\bm{q}) & = -B_{x}(\tau) q_{x}  \notag \\
&\quad  + \left( C_{y}(\tau)q^{2}_{y} + C_{z}(\tau)q^{2}_{z}\right)q_{x} + C_{x}(\tau)q^{3}_{x} , \label{eq: pancake x}\\
y^\BG(\tau,\bm{q}) & = B_{y}(\tau) q_{y} , \label{eq: pancake y}\\
z^\BG(\tau,\bm{q}) & = B_{z}(\tau) q_{z} , \label{eq: pancake z}
\end{align}
where we defined
\begin{align}
B_{x}(\tau) & = - \frac{\partial x^\BG(\tau,\bm{0})}{\partial q_{x}} >0, \\
B_{i}(\tau) & = \frac{\partial a^\BG(\tau,\bm{0})}{\partial q_{x}} >0  \quad \mbox{(for $i=y,z$)}, \\
C_{i}(\tau) & = 
\frac{\partial^{3} x^\BG(\tau,\bm{0})}{\partial q_{x}\partial q_{i}\partial q_{i}}  > 0 \quad \mbox{(for $i=x,y,z$)},
\end{align}
and have adapted sign conventions such that $B_{x}$ becomes positive shortly after the first shell crossing while $B_{i}(\tau)$, $i=y,z$ do not change sign and remain positive during all the course of the  motion.   The positive sign of coefficients $C_{i}(\tau)$  comes from the shell crossing condition, which imposes $J > 0$ before shell crossing time and $J < 0$ just after it in the vicinity of the shell-crossing point, see Eq.~(\ref{eq: detJ pancake}) below. Indeed, we do not consider secondary pancakes which are the result of the merging of previously formed multistream regions. In other words, the multistream region formed shortly after shell crossing is assumed to be compact.

Equations~(\ref{eq: pancake x})--(\ref{eq: pancake z}) are our starting expressions for computing the backreaction from the multistream region in Sec.~\ref{sec: force}. Details on the way these equations are derived, in particular the reason why only linear order in $q_y$ and $q_z$ is necessary in Eqs.~(\ref{eq: pancake y}) and (\ref{eq: pancake z}), are given in Ref.~\cite{2023A&A...678A.168S}.

\subsection{Caustic structure shortly after first shell crossing\label{sec: caustic}}

Before computing the backreaction from the multi-stream region of the flow given in Eqs.~(\ref{eq: pancake x})--(\ref{eq: pancake z}), we investigate the caustic structure of the pancake shortly after the first shell crossing. The caustic structure is given by solving $J(\tau,\bm{q}) = 0$. It provides the boundary between the multi-stream and single-stream regions until second shell-crossing.\footnote{After the second shell crossing takes place, the outermost connected set of curves (in 2D) or surfaces (in 3D) obtained by solving $J(\tau,\bm{q}) = 0$ still provides the boundary between the multi-stream and single-stream regions.}

For the pancake collapse case given by Eqs.~(\ref{eq: pancake x})--(\ref{eq: pancake z}), the caustic structure is determined by solving $\partial x(\tau,\bm{q})/\partial q_{x} = 0$:
\begin{align}
-B_{x}(\tau) + \left\{ C_{y}(\tau)q^{2}_{y} + C_{z}(\tau)q^{2}_{z} \right\} + 3C_{x}(\tau) q^{2}_{x} = 0 , \label{eq: detJ pancake}
\end{align}
where we used Eqs.~(\ref{eq: pancake y}) and (\ref{eq: pancake z}) to remove $y$ and $z$ from the relation. This expression defines the shape of an ellipsoid in Lagrangian space. In particular, we note, that inside the multistream region,
\begin{align}
\left( \frac{q_{y}}{q_{y,{\rm max}}(\tau)}\right)^{2} + \left( \frac{q_{z}}{q_{z,{\rm max}}(\tau)}\right)^{2} \leq 1, \label{eq:ineq}
\end{align}
where $q_{i,{\rm max}}(\tau) \equiv \sqrt{B_{x}(\tau)/C_{i}(\tau)}$ for $i=y$ and $z$.
The quantities $q_{y,{\rm max}}(\tau)$ and $q_{z,{\rm max}}(\tau)$ represent the maximum extent of the Lagrangian caustics of the pancake along $q_y$ and $q_z$ axes.

\begin{figure}
\centering
\includegraphics[width=0.49\textwidth]{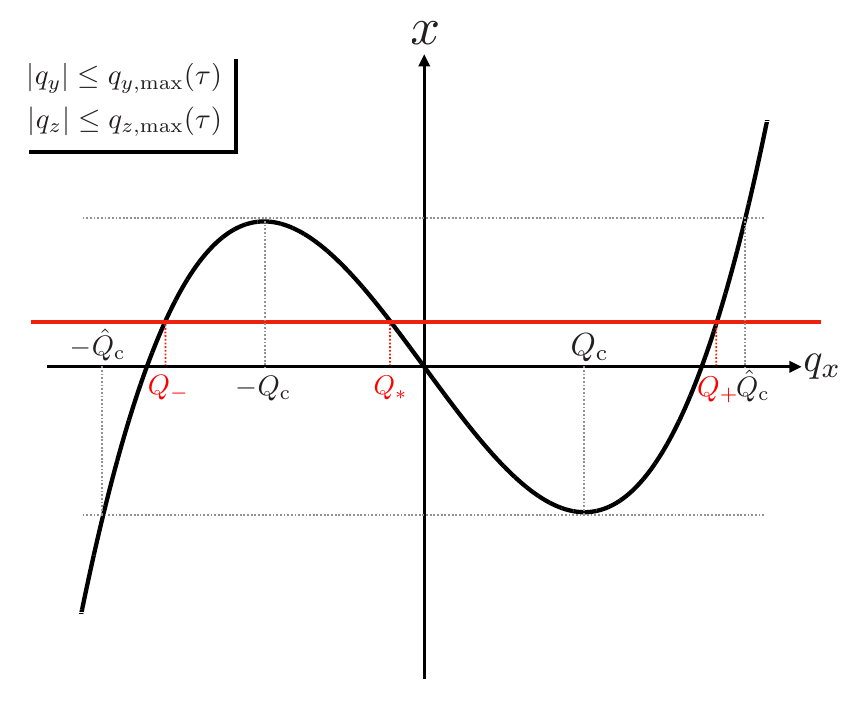}
\caption{Schematic illustration of particle locations in the $q_{x}$-$x$ plane for $(q_{y},q_{z})$-tuples that are located within the caustic.}
\label{fig: schematic}
\end{figure}

Solving Eq.~(\ref{eq: detJ pancake}) in terms of the Lagrangian coordinate~$q_{x}$ and denoting the solution by $Q_{{\rm c}}$, we have
\begin{align}
Q_{{\rm c}}(\tau, q_{y}, q_{z}) = \sqrt{\frac{B_{x}(\tau) -\left( C_{y}(\tau) q_{y}^{2} + C_{z}(\tau) q_{z}^{2} \right) }{3C_{x}(\tau)}
}.  \label{eq: def Q c}
\end{align}
For particles with $(q_y,q_z)$ following inequality (\ref{eq:ineq}), $\pm Q_{{\rm c}}$ denote the Lagrangian position of the upper and lower Eulerian caustic positions along $x$-axis, respectively, in Fig.~\ref{fig: schematic}. As seen on this figure, the width of the multistream region in Lagrangian space is not determined by $Q_{{\rm c}}$, but rather by $\hat{Q}_{{\rm c}} = 2Q_{{\rm c}}$:
\begin{align}
\hat{Q}_{{\rm c}}(\tau, q_{y}, q_{z}) = \sqrt{\frac{4\left\{ B_{x}(\tau) -\left( C_{y}(\tau) q_{y}^{2} + C_{z}(\tau) q_{z}^{2} \right) \right\}}{3C_{x}(\tau)}
}  . \label{eq: def Q c hat}
\end{align}
The expressions of $Q_{{\rm c}}$ and $\hat{Q}_{{\rm c}}$ are analogous to those given for the 1D case in Ref.~\cite{2017MNRAS.470.4858T}, but now account for variations of  $q_{y}$ and $q_{z}$ coordinates. From Eqs.~(\ref{eq: def Q c}) and (\ref{eq: def Q c hat}), we can define the (super-conformal) times $\tau_{{\rm c}}(\bm{q})$ and $\hat{\tau}_{{\rm c}}(\bm{q})$ at which the Lagrangian position $q_{x}$ enters $Q_{{\rm c}}(\tau) < |q_{x}|\leq \hat{Q}_{{\rm c}}(\tau)$ and $|q_{x}|\leq Q_{{\rm c}}(\tau)$, respectively, by 
\begin{align}
Q_{{\rm c}}(\tau_{{\rm c}},q_{y},q_{z}) & \equiv q_{x} , \label{eq: def tau c}\\
\hat{Q}_{{\rm c}}(\hat{\tau}_{{\rm c}},q_{y},q_{z}) & \equiv q_{x} . \label{eq: def tau c hat}
\end{align}
We note that $\tau_{\rm sc}$, $\tau_{{\rm c}}$ and $\hat{\tau}_{{\rm c}}$ are ranked in the following order  $\tau_{\rm sc} < \hat{\tau}_{{\rm c}}(Q,q_{y},q_{z}) < \tau_{{\rm c}}(Q,q_{y},q_{z})$ for $\bm{q} \neq 0$ and are obviously equal otherwise.

Consider the Eulerian coordinate $x$ in the multi-stream region. Given $q_{y}$ and $q_{z}$, the equation $x(\tau,\bm{q}) = x$ has three solutions, denoted by $Q_{*}$ and $Q_{\pm}$ satisfying $Q_{-}<Q_{*}<Q_{+}$ shown in Fig.~\ref{fig: schematic}. Using Eq.~(\ref{eq: pancake x}) and the fact that $x(Q_{-}) = x(Q_{*}) = x(Q_{-})$, the three solutions relate to each other as follows 
\begin{align}
Q_{*} &= \frac{1}{2} \left\{ -Q_{\pm} \pm \sqrt{3\left( \hat{Q}^{2}_{{\rm c}} - Q_{\pm}^{2} \right)} \right\} , \label{eq: Q*}\\
Q_{\pm} &= \frac{1}{2} \left\{ -Q_{*} \pm \sqrt{3\left( \hat{Q}^{2}_{{\rm c}} - Q_{*}^{2} \right)} \right\} , \\
Q_{\pm} &= \frac{1}{2} \left\{ -Q_{\mp} \pm \sqrt{3\left( \hat{Q}^{2}_{{\rm c}} - Q_{\mp}^{2} \right)} \right\} , \label{eq: Qpm}
\end{align}
and are formally equivalent to those derived from the one-dimensional considerations of Refs.~\cite{2015MNRAS.446.2902C,2017MNRAS.470.4858T}.

\subsection{Force in the multi-stream regions \label{sec: force}}

Following~\cite{2023A&A...678A.168S}, we now compute the gravitational force in the multi-stream region of the expanded background flow given in Eqs.~(\ref{eq: pancake x})--(\ref{eq: pancake z}). The key idea is that the size of the pancake along the $x$-axis is much smaller than along the orthogonal directions, suggesting that the structure of the density field inside the pancake and its close vicinity is almost one dimensional. Hence, the three-dimensional problem of solving Poisson equation~(\ref{eq:Poisson2}) can be asymptotically reduced to a one-dimensional problem along the $x$-axis by ignoring the local variations of the density field along the $y$- and $z$-directions.

To facilitate the analysis, we first decompose Poisson equation (\ref{eq:Poisson2}) into two independent equations~\cite{2023A&A...678A.168S}:
\begin{align}
\nabla^{2}_{x}\bar{\Phi} &= - \frac{3}{2}\Omega_{\rm m0}H^{2}_{0}a , \label{eq: bg bar phi}\\
\nabla^{2}_{x}\Phi_{\rm p} &= \frac{3}{2}\Omega_{\rm m0}H^{2}_{0}a\left( 1 + \delta(\bm{x}) \right) , \label{eq: phi p}
\end{align}
where $\Phi = \bar{\Phi} + \Phi_{\rm p}$.
The first and second terms of the right-hand side of this equation are, respectively, the gravitational potential coming from the total density $ \bar{\rho} (1 + \delta)$ and that coming from the negative background density $-\bar{\rho}$ as a counter term. We also define the corresponding forces by $\bar{\bm{F}} = -\bm{\nabla}_{x}\bar{\Phi}(\bm{x})$ and $\bm{F} = -\bm{\nabla}_{x}\Phi_{\rm p}(\bm{x})$. Eq.~(\ref{eq: bg bar phi}) can be readily solved as
\begin{align}
\bar{\bm{F}}(\tau, \bm{x}) & = \frac{3}{2}\Omega_{\rm m0}H^{2}_{0}a \frac{1}{d} \bm{x} ,  \label{eq: bg force}
\end{align}
with $d = 2$ and $3$ for the two- and three-dimensional cases, respectively. We note that Ref.~\cite{2023A&A...678A.168S} found that this counter term does not contribute very much to the $x$-direction of the force in the multi-stream regions.

Importantly, as shown in Ref.~\cite{2023A&A...678A.168S}, the gravitational force along $y$- and $z$-directions does not change qualitatively before and after shell crossing and preserves the single-stream nature of the flow, suggesting that we can use the pre-collapse prediction, e.g., LPT solutions, for the motion along $y$- and $z$-axes, without adding the backreaction from the multi-stream region, as long as the perturbative series converges. Hence, what we need to care about, concerning the multi-stream backreaction, is the $x$-directional motion. Hereafter, we focus on the $x$-direction of the force and solve Eq.~(\ref{eq: phi p}). Assuming that the collapsing structure is a pancake, given fixed $(q_{y}, q_{z})$, we approximate Eq.~(\ref{eq: phi p}) by~\citep[see Ref.~][]{2023A&A...678A.168S}
\begin{align}
\frac{{\rm d}^{2}\Phi_{\rm p}(\tau,\bm{x}(\tau,\bm{q}))}{{\rm d}x^{2}}
& \simeq \frac{1}{B_{y}(\tau)B_{z}(\tau)}\frac{3}{2}\Omega_{\rm m0}H^{2}_{0}a\left| \frac{\partial x}{\partial q_{x}} \right|^{-1} . \label{eq: 1D Poisson eq}
\end{align}
Since this expression is now a purely one-dimensional Poisson equation, we exploit the Green function method of the one-dimensional problem to solve this equation, exactly as in Ref.~\cite{2017MNRAS.470.4858T} \citep[see also][for similar derivations]{2021MNRAS.505L..90R}. Then, the $x$-component of the force at $x=x(\tau,\bm{q})$ is given by (see Appendix~\ref{sec: app Fx} for the detailed derivations)
\begin{align}
F_{x}(\tau,\bm{q}) 
& = -\frac{3}{2}\Omega_{\rm m0}H^{2}_{0}
\frac{a}{B_{y}(\tau)B_{z}(\tau)} 
\mathcal{F}(\tau, \bm{q}).
\label{eq: Fx}
\end{align}
In the above, we defined
\begin{align}
\mathcal{F}(\tau, \bm{q}) = 
\left[-2 + \beta(\tau,q_{y},q_{z}) \right]q_{x} - C_{x}(\tau) q_{x}^{3}
, \label{eq: cal F in}
\end{align}
for $|q_{x}| \leq Q_{{\rm c}}(\tau,q_{y},q_{z})$, and
\begin{align}
\mathcal{F}(\tau, \bm{q}) 
&= 
\left[1 + \beta(\tau,q_{y},q_{z}) \right] q_{x} - C_{x}(\tau) q_{x}^{3}
\notag \\
&
- {\rm sgn}(q_{x})\sqrt{3}\sqrt{ \hat{Q}_{{\rm c}}^{2}(\tau,q_{y},q_{z})-q_{x}^{2} }
, \label{eq: cal F out}
\end{align}
for $Q_{{\rm c}}(\tau,q_{y},q_{z})<|q_{x}| \leq \hat{Q}_{{\rm c}}(\tau,q_{y},q_{z})$, 
with
\begin{align}
  \beta(\tau,q_{y},q_{z}) = B_{x}(\tau) - \left( C_{y}(\tau) q_{y}^{2} + C_{z}(\tau) q_{z}^{2} \right).
\end{align}
The above expressions generalize the one-dimensional results of Refs.~\cite{2017MNRAS.470.4858T,2021MNRAS.505L..90R} [see their Eq.~(42) and Eqs.~(40)-(42), respectively] to the two- and three-dimensional cases, for fixed Lagrangian coordinates $(q_{y},q_{z})$ that directly relate to the Eulerian coordinates $(y,z)$ via Eqs.~(\ref{eq: pancake y}) and (\ref{eq: pancake z}).

\subsection{Time dependence shortly after shell crossing\label{sec: x exp}}

Equation~(\ref{eq: Fx}) together with Eqs.~(\ref{eq: cal F in}) and (\ref{eq: cal F out}) does not assume explicit forms for the time dependence of the background flow $x^\BG(\tau,\bm{q})$ in Eqs.~(\ref{eq: pancake x})--(\ref{eq: pancake z}). In this sense, the expressions given in the previous subsection are generic formula for the $x$-component of the force inside an axisymmetric pancake, although strictly (asymptotically) valid only at times very close to shell crossing. To perform the calculations of the correction to the motion, the explicit time dependence is however presently required because we are now integrating Eqs.~(\ref{eq: x formal}) and (\ref{eq: u formal}) with respect to (super-conformal) time.

To do so, since we are interested in the fluid motion shortly after the first shell-crossing time $\tau_{\rm sc}$, we further expand the expression $\bm{x}^\BG(\tau, \bm{q})$ in Eq.~(\ref{eq: expand x}) at linear order in $\tau-\tau_{\rm sc}$:
\begin{align}
\bm{x}^\BG(\tau,\bm{q}) \simeq \bm{x}^\BG(\tau_{\rm sc},\bm{q}) + \bm{u}^\BG(\tau_{\rm sc},\bm{q}) \left( \tau - \tau_{\rm sc} \right) , \label{eq: ballistic0}
\end{align}
where we use $\bm{u}^\BG = {\rm d}\bm{x}^\BG/{\rm d}\tau$. Note that this ballistic approximation justifies {\em a posteriori} why only linear terms in $q_y$ and $q_z$ are kept in Eqs.~(\ref{eq: pancake y}) and (\ref{eq: pancake z}), as detailed in Ref.~\cite{2023A&A...678A.168S}.

In this simplified framework, the time-dependent functions $B_{i}$ and $C_{i}$ are given by
\begin{align}
B_{x}(\tau) & = B^{(1)}_{x}(\tau - \tau_{\rm sc}) , \\
B_{a}(\tau) & = B^{(0)}_{a} - B^{(1)}_{a}(\tau - \tau_{\rm sc}), \\
C_{b}(\tau) & = C^{(0)}_{b} + C^{(1)}_{b}(\tau - \tau_{\rm sc}),
\end{align}
with the constant coefficients being explicitly given by
\begin{align}
B^{(0)}_{i} & = \frac{\partial a^\BG(\tau_{\rm sc},\bm{0})}{\partial q_{i}} , \\
B^{(1)}_{j} &= - \frac{\partial u^\BG_{b}(\tau_{\rm sc},\bm{0})}{\partial q_{j}}, \\
C^{(0)}_{x} & = \frac{1}{6}\frac{\partial^{3} x^\BG(\tau_{\rm sc},\bm{0})}{\partial q_{x}^{3}} , \\
C^{(1)}_{x}  &= \frac{1}{6}\frac{\partial^{3} u^\BG_{x}(\tau_{\rm sc},\bm{0})}{\partial q_{x}^{3}} , \\
C^{(0)}_{i} & = \frac{1}{2}\frac{\partial^{3} a^\BG(\tau_{\rm sc},\bm{0})}{\partial q_{x}\partial q_{i}^{2}}, \\
C^{(1)}_{i} &= \frac{1}{2}\frac{\partial^{3} u^\BG_{x}(\tau_{\rm sc},\bm{0})}{\partial q_{x}\partial q_{i}^{2}} ,
\end{align}
where $i=y,z$ and $j=x,y,z$. In the above, we do not apply Einstein summation convention for repeated indices. Using these explicit forms, we calculate the quantities relevant to the caustics, $Q_{{\rm c}}$ and $\hat{Q}_{{\rm c}}$, as follows:
\begin{align}
\hat{Q}_{{\rm c}}(\tau,q_{y},q_{z}) 
& = 2Q_{{\rm c}}(\tau,q_{y},q_{z}) \notag \\
& \simeq 2\sqrt{-\kappa^{(0)} (q_{y},q_{z}) + \frac{2(\tau-\tau_{\rm sc})}{\kappa^{(1)} (q_{y},q_{z})}} . \label{eq: hat Q approx}
\end{align}
Here the functions $\kappa^{(0)} (q_{y},q_{z})$ and $\kappa^{(1)} (q_{y},q_{z})$ are defined by
\begin{align}
\kappa^{(0)} (q_{y},q_{z}) & = \frac{1}{3C^{(0)}_{x}}\left[  C^{(0)}_{y} q_{y}^{2} + C^{(0)}_{z} q_{z}^{2} \right] , \label{eq: kappa 0}\\
\frac{1}{\kappa^{(1)} (q_{y},q_{z})} & = \frac{1}{6\left( C^{(0)}_{x} \right)^{2}}
\Biggl[
B^{(1)}_{x}C^{(0)}_{x}
\notag \\
&
+ q_{y}^{2} (C^{(1)}_{x}C^{(0)}_{y}-C^{(0)}_{x}C^{(1)}_{y})
\notag \\
&
+ q_{z}^{2} (C^{(1)}_{x}C^{(0)}_{z}-C^{(0)}_{x}C^{(1)}_{z})
\Biggr] . \label{eq: kappa 1}
\end{align}
Note that, in the subsequent calculations, the full expression of $\kappa^{(1)} (q_{y},q_{z})$ as defined in Eq.~(\ref{eq: kappa 1}) will be used.\footnote{However, setting $q_y=q_z=0$ in Eq.~(\ref{eq: kappa 1}) would be equally valid from the asymptotic point of view, because the extension of the multistream region scales like $\sqrt{\tau -\tau_{\rm c}}$ in Lagrangian space, $q_{x,{\rm max}}(\tau)={\hat Q}_{\rm c}(\tau,0,0)\sim q_{y,{\rm max}}(\tau) \sim q_{z,{\rm max}}(\tau) \sim (\tau-\tau_{\rm c})^{1/2}$, see Ref.~\cite{2023A&A...678A.168S}.}
Using Eq.~(\ref{eq: hat Q approx}), the quantities $\tau_{{\rm c}}$ and $\hat{\tau}_{{\rm c}}$ defined in Eqs.~(\ref{eq: def tau c}) and (\ref{eq: def tau c hat}), respectively, are calculated as
\begin{align}
\hat{\tau}_{{\rm c}}(\bm{q}) & \simeq \tau_{\rm sc} + \frac{1}{2}\kappa^{(0)} (q_{y},q_{z}) \kappa^{(1)} (q_{y},q_{z}) + \frac{1}{8}\kappa^{(1)} (q_{y},q_{z})q_{x}^{2} , \label{eq: tau c hat}\\
\tau_{{\rm c}}(\bm{q}) & \simeq \tau_{\rm sc} + \frac{1}{2} \kappa^{(0)} (q_{y},q_{z}) \kappa^{(1)} (q_{y},q_{z}) + \frac{1}{2}\kappa^{(1)} (q_{y},q_{z})q_{x}^{2} . \label{eq: tau c}
\end{align}
Finally, from Eqs.~(\ref{eq: Fx}), (\ref{eq: cal F in}) and (\ref{eq: cal F out}), we can estimate the $x$-component of the force up to the linear level in $\tau-\tau_{\rm sc}$:
\begin{align}
F_{x}(\tau,\bm{q}) 
  & \simeq -\frac{3}{2}\Omega_{\rm m0}H^{2}_{0} \frac{a(\tau_{\rm sc})}{B^{(0)}_{y} B^{(0)}_{z}} \notag \\
 & \times \left[ \mathcal{A}^{(0)}(\bm{q}) + \mathcal{A}^{(1)}(\bm{q})  (\tau-\tau_{\rm sc}) \right] , \\
& \equiv F^{\rm in}_{x}(\tau,\bm{q}) ,\label{eq:fxin}
\end{align}
for $|q_{x}| \leq Q_{{\rm c}}(\tau,q_{y},q_{z})$, and
\begin{align}
F_{x}(\tau,\bm{q})
& \simeq -\frac{3}{2}\Omega_{\rm m0}H^{2}_{0} \frac{a(\tau_{\rm sc})}{B^{(0)}_{y} B^{(0)}_{z}}
\notag \\
& \times 
\Bigl[
\mathcal{B}^{(0)}(\bm{q})
+ \mathcal{B}^{(1)}(\bm{q})  (\tau-\tau_{\rm sc})
\notag \\
& 
- {\rm sgn}(q_{x})\sqrt{3}\sqrt{ \hat{Q}_{{\rm c}}^{2}(\tau,q_{y},q_{z})-q_{x}^{2} }
\Bigr] , \\
& \equiv F^{\rm out}_{x}(\tau,\bm{q}) , \label{eq:fxout}
\end{align}
for $Q_{{\rm c}}(\tau,q_{y},q_{z})<|q_{x}| \leq \hat{Q}_{{\rm c}}(\tau,q_{y},q_{z})$.
The third order polynomials $\mathcal{A}^{(0),(1)}(\bm{q})$ and $\mathcal{B}^{(0),(1)}(\bm{q})$ are explicitly given in Eqs.~(\ref{eq: calF in 0})--(\ref{eq: calF out 1}) in Appendix~\ref{sec: app Fx}.

It is important to note that the terms linear in $\tau-\tau_{\rm c}$ in  equations (\ref{eq:fxin}) and (\ref{eq:fxout}) are only approximate.  To get these contributions correctly and in a way consistent with catastrophe theory, we would need to go beyond the ballistic approximation in Eq.~(\ref{eq: ballistic0}) and beyond third order in the Taylor expansion (\ref{eq: expand x}) by employing an iterative procedure. While we provided them here for completeness because they are useful to check {\em a posteriori} the validity of the PCPT approach when computing the time integrals (\ref{eq: def Delta u out}),  (\ref{eq: def Delta x out}), (\ref{eq: def Delta u in 2}) and (\ref{eq: def Delta x in 2}) below, these next to leading order terms in time will be neglected in next section. 

\subsection{PCPT: corrections from the multi-stream flow\label{sec: formulation}}

We are now in a position to derive the corrections to the background motion from the multi-stream flow in the framework of the perturbative treatments given in Eqs.~(\ref{eq: expand x}) and (\ref{eq: ballistic0}). Since the expression of the force depends in a non trivial way on the Lagrangian position as pictured by Eqs.~(\ref{eq: Fx}),  (\ref{eq: cal F in}) and (\ref{eq: cal F out}), in order to compute the corrections, we need to divide the domain of the integrals in Eqs.~(\ref{eq: Delta x formal}) and (\ref{eq: Delta u formal}) into three pieces \citep[see also Sec.3.3 in][]{2017MNRAS.470.4858T}.

We first consider a mass element inside the outer part of the multi-stream region in Lagrangian space, with $Q_{{\rm c}}(\tau,q_{y},q_{z}) < |q_{x}| \leq \hat{Q}_{{\rm c}}(\tau,q_{y},q_{z})$, or, equivalently, verifying $\hat{\tau}_{{\rm c}}(\bm{q}) \leq \tau < \tau_{{\rm c}}(\bm{q})$. In this case, we have
\begin{align}
x(\tau,\bm{q}) &=
x^\BG(\hat{\tau}_{{\rm c}}(\bm{q}),\bm{q}) 
+ u^\BG_{x}(\hat{\tau}_{{\rm c}}(\bm{q}),\bm{q}) \left[ \tau - \hat{\tau}_{{\rm c}}(\bm{q}) \right]
\notag \\
& 
+ \delta x^{{\rm out}}(\tau,\bm{q}) , \\
u_{x}(\tau,\bm{q}) &= u^\BG_{x}(\hat{\tau}_{{\rm c}}(\bm{q}),\bm{q}) + \delta u^{{\rm out}}_{x}(\tau,\bm{q}) , 
\end{align}
with
\begin{align}
\delta u^{{\rm out}}_{x}(\tau,\bm{q}) & = \int^{\tau}_{\hat{\tau}_{{\rm c}}(\bm{q})}{\rm d}\tau'\, F^{\rm out}_{x}(\tau', \bm{q}) , \label{eq: def Delta u out} \\
\delta x^{{\rm out}} (\tau,\bm{q}) & = \int^{\tau}_{\hat{\tau}_{{\rm c}}(\bm{q})}{\rm d}\tau'\, \delta u^{{\rm out}}_{x}(\tau', \bm{q}) . \label{eq: def Delta x out}
\end{align}

Second, we consider a mass element entering subsequently the inner part of the multi-stream region in Lagrangian space, with $|q_{x}| \leq Q_{{\rm c}}(\tau,q_{y},q_{z})$ or, equivalently, verifying $\tau \geq \tau_{{\rm c}}(\bm{q})$. In this case, we have 
\begin{align}
x(\tau,\bm{q}) &= 
x^\BG(\hat{\tau}_{{\rm c}}(\bm{q}),\bm{q}) 
+ u^\BG_{x}(\hat{\tau}_{{\rm c}}(\bm{q}),\bm{q}) \left[ \tau - \hat{\tau}_{{\rm c}}(\bm{q}) \right]
\notag \\
& 
+ \delta x^{{\rm in}}(\tau,\bm{q}) , \\
u_{x}(\tau,\bm{q}) &= u^\BG_{x}(\hat{\tau}_{{\rm c}}(\bm{q}),\bm{q}) + \delta u^{{\rm in}}_{x}(\tau,\bm{q}) , 
\end{align}
with
\begin{align}
\delta u^{{\rm in}}_{x}(\tau,\bm{q}) & =
\int^{\tau_{{\rm c}}(\bm{q})}_{\hat{\tau}_{{\rm c}}(\bm{q})}{\rm d}\tau'\, F^{\rm out}_{x}(\tau', \bm{q})
\notag \\
& \quad
+ \int^{\tau}_{\tau_{{\rm c}}(\bm{q})}{\rm d}\tau'\, F^{\rm in}_{x}(\tau', \bm{q}) ,
\label{eq: def Delta u in 2}\\
\delta x^{{\rm in}} (\tau,\bm{q}) 
& = \int^{\tau_{{\rm c}}(\bm{q})}_{\hat{\tau}_{{\rm c}}(\bm{q})}{\rm d}\tau'\, \delta u^{{\rm out}}_{x}(\tau', \bm{q})
\notag \\
& \quad
+ \int^{\tau}_{\tau_{{\rm c}}(\bm{q})}{\rm d}\tau'\, \delta u^{{\rm in}}_{x}(\tau', \bm{q}) . \label{eq: def Delta x in 2}
\end{align}
Finally, for fluid particles in the single-stream regime, $\hat{Q}_{{\rm c}}(\tau,q_{y},q_{z})<|q_{x}|$, we simply use the background flow: $x(\tau,\bm{q}) = x^\BG(\tau,\bm{q})$ and $u_{x}(\tau,\bm{q}) = u^\BG_{x}(\tau,\bm{q})$.

After a straightforward, though lengthy, calculation, we obtain the corrections to the motion $\delta x^{\rm in/out}$ and $\delta u^{\rm in/out}_{x}$. Leaving important details of the derivations to Appendix~\ref{sec: app derivations}, we simply show the final expressions here:
\begin{align}
\delta x^{{\rm out}}(\tau,\bm{q}) 
& = 
-\frac{3}{2}\Omega_{\rm m0}H^{2}_{0}\frac{a(\tau_{\rm sc})}{B^{(0)}_{y} B^{(0)}_{z}}
\times 
\Biggl[ 
\frac{1}{2}\mathcal{B}^{(0)} \left( \tau - \hat{\tau}_{{\rm c}} \right)^{2}
\notag \\
& 
- {\rm sgn}(q_{x})\frac{\sqrt{3}}{240}\left(\kappa^{(1)}\right)^{2} \left( \hat{Q}_{{\rm c}}^{2}(\tau)-q_{x}^{2} \right)^{5/2}
\Biggr] , \label{eq: Delta x out} \\
\delta u^{{\rm out}}_{x}(\tau,\bm{q}) 
& = -\frac{3}{2}\Omega_{\rm m0}H^{2}_{0} \frac{a(\tau_{\rm sc})}{B^{(0)}_{y} B^{(0)}_{z}}
\Biggl[ 
\mathcal{B}^{(0)} \left( \tau - \hat{\tau}_{{\rm c}} \right)
\notag \\
&
- {\rm sgn}(q_{x})\frac{\sqrt{3}}{12}\kappa^{(1)}  \left( \hat{Q}_{{\rm c}}^{2}(\tau)-q_{x}^{2}\right)^{3/2}
\Biggr] , \label{eq: app Delta u out}
\end{align}
\begin{align}
\delta x^{{\rm in}}(\tau,\bm{q}) 
& =
-\frac{3}{2}\Omega_{\rm m0}H^{2}_{0}\frac{a(\tau_{\rm sc})}{B^{(0)}_{y} B^{(0)}_{z}} \notag \\
& \times \Biggl[ 
\mathcal{B}^{(0)}\frac{9}{128}\left(\kappa^{(1)}\right)^{2}q_{x}^{4} 
- \frac{9}{80} \left(\kappa^{(1)}\right)^{2} q_{x}^{5}
\notag \\
&
\quad + \Bigl\{
\mathcal{B}^{(0)} \frac{3}{8}\kappa^{(1)} q_{x}^{2}
- \frac{3}{4}\kappa^{(1)}  q_{x}^{3}
\Bigr\} \left( \tau - \tau_{{\rm c}} \right) \notag \\
& \quad + \frac{1}{2}\mathcal{A}^{(0)} (\tau - \tau_{{\rm c}})^{2}
\Biggr] , \label{eq: Delta x in}\\
\delta u^{{\rm in}}_{x}(\tau,\bm{q}) 
& =
-\frac{3}{2}\Omega_{\rm m0}H^{2}_{0}\frac{a(\tau_{\rm sc})}{B^{(0)}_{y} B^{(0)}_{z}} \notag \\
& \times \Biggl[
\mathcal{B}^{(0)} \frac{3}{8}\kappa^{(1)} q_{x}^{2}
- \frac{3}{4}\kappa^{(1)}  q_{x}^{3}
+ \mathcal{A}^{(0)} (\tau - \tau_{{\rm c}}) \Biggr], \label{eq: Delta u in} 
\end{align}
where the $\bm{q}$ dependence of functions $\kappa^{(1)}(q_x,q_y)$, ${\hat Q}_{\rm c}(\tau, q_x,q_y)$ and third order polynomials $\mathcal{A}^{(0)}(\bm{q})$, $\mathcal{B}^{(0)}(\bm{q})$ (see Appendix~\ref{sec: app Fx}) has been kept implicit for simplicity. In the expressions above, contributions of up to 7th order in $q_x$ are accounted for, but these are only meant to connect the three Lagrangian domains in order to produce a fully continuous solution.\footnote{Note again that we keep the expression of  $\kappa^{(1)}$ as given by Eq.~(\ref{eq: kappa 1}) without Taylor expanding it.} The solutions are actually asymptotically valid only up to third order in the Lagrangian coordinate, and linear and quadratic order in time, $\tau-\tau_{\rm c}$, for the velocity and the position, respectively.

These expressions represent the main analytical results of this paper and provide us with a perturbative way to describe the post-collapse motion of a CDM pancake in three-dimensional space. They can be regarded as a generalization to 3D of the one-dimensional PCPT formalism developed in Refs.~\cite{2015MNRAS.446.2902C,2017MNRAS.470.4858T}. While the above expressions hold for $\Lambda$CDM cosmology, we will test the performances of these analytical results in Sec.~\ref{sec: comparison} for the Einstein-de Sitter case.

To conclude this technical section, let us summarize the key steps that allowed us to derive our main results:
  \begin{enumerate}
  \item{\em Reduction to a 1D problem:} the PCPT framework reduces the three-dimensional problem to a one-dimensional problem along the collapse direction (in the present case, the $x$ direction) by exploiting the fact that, shortly after shell-crossing, the multistream region is thin along $x$ but wide along the transverse directions ($y, z$). This is a generic property of collapse except when the flow is locally perfectly spherical or cylindrical, which is, strictly speaking from the mathematical point of view, a zero probability event. This asymptotic property allows a controlled dimensional reduction of the three-dimensional Poisson equation to Eq.~(\ref{eq: 1D Poisson eq}), which is a one-dimensional Poisson equation. 
\item{\em Cubic-order spatial expansion:} the PCPT framework expands the displacement field in terms of the Lagrangian position $\bm{q}$ up to cubic order about the shell-crossing point, which is sufficient to capture the topology that dictates the post-collapse motion along the $x$ direction. The cubic order expansion is crucial for accurately modeling the three-stream flow and the gravitational backreaction. 
\item{\em Ballistic approximation:} once the first shell crossing occurs, the PCPT framework assumes that the backreactions from the multi-stream region are captured by employing a ballistic approximation for the background flow (see Eq.~\ref{eq: ballistic0}). This assumption is mathematically justified in the asymptotic regime shortly after collapse. It also suppresses, to some extent, the possibly non-convergent behavior of the LPT series after the first shell crossing.
\end{enumerate}

\section{Comparison to simulations\label{sec: results}}

In the previous section, we developed a perturbative treatment to describe the post-collapse motion of pancake collapse in three-dimensional space. We now examine the validity of our formalism by comparing it with simulations. In doing so, we need to specify the \textit{background flow} (to be precise, functional form of $\bm{x}^\BG(\tau,\bm{q})$ and $\bm{u}^\BG(\tau,\bm{q})$), for which we use high-order LPT solutions, as detailed in Sec.~\ref{sec: LPT}. We also need to specify the initial conditions of the system so as to satisfy the axisymmetic pancake collapse conditions. As a representative set of initial conditions, we investigate systems seeded with  two- or three-sine waves, as introduced in Sec.~\ref{sec: IC}. In Sec.~\ref{sec: comparison}, we compare our analytical results with measurements in Vlasov-Poisson simulations performed with the public code \texttt{ColDICE}~\cite{2016JCoPh.321..644S}.

\subsection{Lagrangian perturbation theory \label{sec: LPT}}

In LPT, we systematically expand the displacement field as~\cite{1989RvMP...61..185S,1992ApJ...394L...5B,1992MNRAS.254..729B,1993MNRAS.264..375B,1995A&A...296..575B,1994ApJ...427...51B}
\begin{align}
  \bm{\psi}(\tau,\bm{q}) = \sum_{n=1}^{\infty} \bm{\psi}^{(n)}(\tau,\bm{q}).
  \label{eq:exp}
\end{align}
Assuming that the fastest growing modes dominate~\cite{2002PhR...367....1B}, Eq.~(\ref{eq:exp}) is well approximated by
\begin{align}
\bm{\psi}(\tau,\bm{q}) \simeq \sum_{n=1}^{\infty} \left( D_{+}^n(\tau) \right)^{n} \bm{\psi}^{(n)}(\bm{q}), \label{eq:LPT}
\end{align}
with $D_{+}$ being the linear growth factor. Thanks to the recurrence relations of LPT, e.g., Refs.~\cite{2012JCAP...12..004R,2014JFM...749..404Z,2015MNRAS.452.1421R,2015PhRvD..92b3534M}, we can obtain the high-order solutions $\bm{\psi}^{(n)}$, which in practice are known to converge with $n$ at least until shell-crossing \cite{2014JFM...749..404Z,Rampf:2017jan,Schmidt:2020ovm,2021MNRAS.501L..71R,2022PhRvF...7j4610R,Rampf:2024uvj}.

Substituting Eq.~(\ref{eq:LPT}) into Eqs.~(\ref{eq:x-Psi}) and (\ref{eq:u-Psi}), we express the Eulerian position and velocity in terms of the $n$th-order LPT solutions:
\begin{align}
\bm{x}^{n{\rm LPT}}(\tau,\bm{q}) & = \bm{q} + \sum_{m=1}^{n}\left(D_{+}(\tau)\right)^{m} \bm{\psi}^{(m)}(\bm{q}), \label{eq: LPT x}\\
\bm{u}^{n{\rm LPT}}(\tau,\bm{q}) & = a^{2}Hf\sum_{m=1}^{n}m\left(D_{+}(\tau)\right)^{m}\bm{\psi}^{(m)}(\bm{q}) , \label{eq: LPT u}
\end{align}
where $f\equiv {\rm d} \ln D_+/{\rm d} \ln a$.
With the $n$th-order LPT solution (\ref{eq: LPT x}), the Jacobian matrix is given by:
\begin{align}
J^{n{\rm LPT}}_{ij}(\tau,\bm{q})
&= \delta_{ij} + \sum^{n}_{m=1}\left( D_{+}(\tau) \right)^{m} \frac{\partial \psi^{(m)}_{i}(\bm{q})}{\partial q_{j}}.
\end{align}
The time $\tau_{\rm sc}$ of first shell-crossing taking place at the Lagrangian position $\bm{q}=\bm{q}_{\rm sc}$ is estimated in LPT by the first occurrence of $\det\left(J^{n{\rm LPT}}_{ij}(\tau_{\rm sc},\bm{q}_{\rm sc})\right) = 0$. As a background flow $\bm{x}^\BG(\tau,\bm{q})$ in Eq.~(\ref{eq: expand x}), we will choose the high-order LPT solution $\bm{x}^{n{\rm LPT}}(\tau,\bm{q})$ with $n=15$ (see Appendix~\ref{sec: app LPT dependence} for the effects of changing the LPT order $n$ on the solution), then correct this background LPT motion by taking into account the multi-stream backreaction.

\subsection{Initial conditions and simulations \label{sec: IC}}
\begin{table*}
\setlength{\tabcolsep}{9pt}
\centering
\caption{Parameters of the runs performed with the public Vlasov code \texttt{ColDICE}~\cite{2016JCoPh.321..644S} along with the relevant crossing times predicted by standard 15LPT and measured in the simulations. 
The first column indicates the designation of the runs. 
The second and third columns display the relative amplitudes of the initial sine waves, namely, $\epsilon_{\rm 2D} = \epsilon_{y}/\epsilon_{x}$ (2D) and $\bm{\epsilon}_{\rm 3D} = (\epsilon_{y}/\epsilon_{x}, \epsilon_{z}/\epsilon_{x})$ (3D), and the value of $\epsilon_{x}$ itself, respectively. 
The fourth and fifth columns show the shell-crossing times from standard 15LPT predictions along the $x$ and $y$-directions, respectively, while the three next ones provide estimates of these times in the \texttt{ColDICE} simulations. These latter times are expected to be accurate at least up to the $10^{-3}$ level.  In the numerical simulation case, we also provide the estimated time of second shell crossing along $x$ axis (sixth column), which takes place before shell crossing along $y$ axis in the Q1D cases. Other details of the runs can be found in the main text and Table~1 of Refs.~\cite{2021A&A...647A..66C,2022A&A...664A...3S,2023A&A...678A.168S}.}
\begin{tabular}{ccccccccc}
\hline\hline
Designation & $\epsilon_{\rm 2D}$ or $\bm{\epsilon}_{\rm 3D}$ & $\epsilon_{x}$ & $a^{\rm 15LPT}_{{\rm sc},x}$ & $a^{\rm sim}_{{\rm sc},x}$ & $a^{\rm sim}_{{\rm sc},x,2}$ & $a^{\rm 15LPT}_{{\rm sc},y}$ & $a^{\rm sim}_{{\rm sc},y}$\\
\hline
{\it Quasi 1D}\\
Q1D-2SIN & 1/6 & -18 & 0.05279 & 0.05285 & 0.1043 & 0.1057 & 0.1383 \\
Q1D-3SIN & (1/6, 1/8) & -24 & 0.03814 & 0.03832 &  0.0693 &0.07455 & 0.0889\\
\hline
{\it Anisotropic}\\
ANI-2SIN & 2/3 & -18 & 0.04540 & 0.04545 & 0.0590 & 0.05409 & 0.0538\\
ANI-3SIN & (3/4, 1/2) & -24 &  0.02912 & 0.02919 & 0.0360 & 0.03233 & 0.0317 \\
\hline\hline
\end{tabular}
\label{tab: ICs}
\end{table*}

To test PCPT, we consider initial conditions seeded by two or three crossed sine waves in a periodic box $[-L/2, L/2[$~\cite{1991ApJ...382..377M,1995ApJ...441...10M} \citep[see also our previous works][]{2018PhRvL.121x1302S,2022A&A...664A...3S,2023A&A...678A.168S}. The initial displacement field at the starting time $\tau_{\rm ini}$ is given by
\begin{align}
\psi^{\rm ini}_{i}(\bm{q},\tau_{\rm ini}) = \frac{L}{2\pi} D_{+}(\tau_{\rm ini}) \epsilon_{i} \sin\left( \frac{2\pi}{L} q_{i}\right) ,
\end{align}
with $\epsilon_{i}<0$ and $|\epsilon_{x}| > |\epsilon_{y}| > |\epsilon_{z}|$. The initial density field presents a small density peak at the origin in these initial conditions. As the first shell crossing takes place along the $x$ direction, this initial setup is directly compatible with the underlying assumptions of previous section and provides a good test case of PCPT. In the subsequent investigations, we consider the Einstein-de Sitter Universe, in which we have $D_{+} = a$, $f = 1$ and $\tau=-2/\sqrt{a}$. The ratios of the parameters $\epsilon_{\rm 2D} = \epsilon_{y}/\epsilon_{x}$ for two sine waves and $\bm{\epsilon}_{\rm 3D} = (\epsilon_{y}/\epsilon_{x},\epsilon_{z}/\epsilon_{x})$ for three sine waves control the dynamics. We consider two configurations: the quasi one-dimensional (Q1D) case, with $|\epsilon_{x}|\gg|\epsilon_{y,z}|$, the anisotropic (ANI) case, with $|\epsilon_{x}|>|\epsilon_{y}|>|\epsilon_{z}|$ but of the same order, as summarized in Table~\ref{tab: ICs}.

Since the initial conditions are given by crossed sine waves, the LPT recurrence relation is simplified and allows us to easily derive high-order LPT solutions in terms of trigonometric polynomials~\cite{2018PhRvL.121x1302S,2022A&A...664A...3S,2023A&A...678A.168S}. We use 15LPT (order $n=15$) as a primary choice for the background flow, which ensures high fidelity in describing dark mater fluid motion even for times close to the first shell-crossing, but we will also discuss the dependence of the results on LPT order later.

Our analytical predictions will be compared with \texttt{ColDICE}~\cite{2016JCoPh.321..644S} simulations performed by~\cite{2021A&A...647A..66C} and \cite{2022A&A...664A...3S}. The Vlasov solver \texttt{ColDICE} directly follows the evolution of a $d$-dimensional phase-space sheet moving in $2d$-dimensional phase space with an adaptive tessellation of triangles or tetrahedra respectively for $d=2$ or $3$, and solves Poisson equation on a mesh of fixed resolution.

Since the technical details on the simulations used in the present work, in particular critical parameters such as those controlling mesh and tessellation resolution, have already been given in Refs.~\cite{2021A&A...647A..66C,2022A&A...664A...3S,2023A&A...678A.168S}, we do not repeat all these pieces of information here. Instead, we simply provide a summary of the simulations setups in Table~\ref{tab: ICs} along with values of various crossing times measured in the runs and predicted by LPT, which will be relevant in the subsequent discussions.

\subsection{Numerical results \label{sec: comparison}}

We are now ready to discuss the validity of the analytical predictions based on the post-collapse treatment by comparing them with the results of the Vlasov-Poisson simulations. We use the following units in the subsequent presentation: a box size $L=1$ and a Hubble parameter at present time $H_{0} = 1$ for the dimensions of length and time, respectively.

\subsubsection{Lagrangian phase-space slices}

\begin{figure*}
\centering
\includegraphics[width=0.97\textwidth]{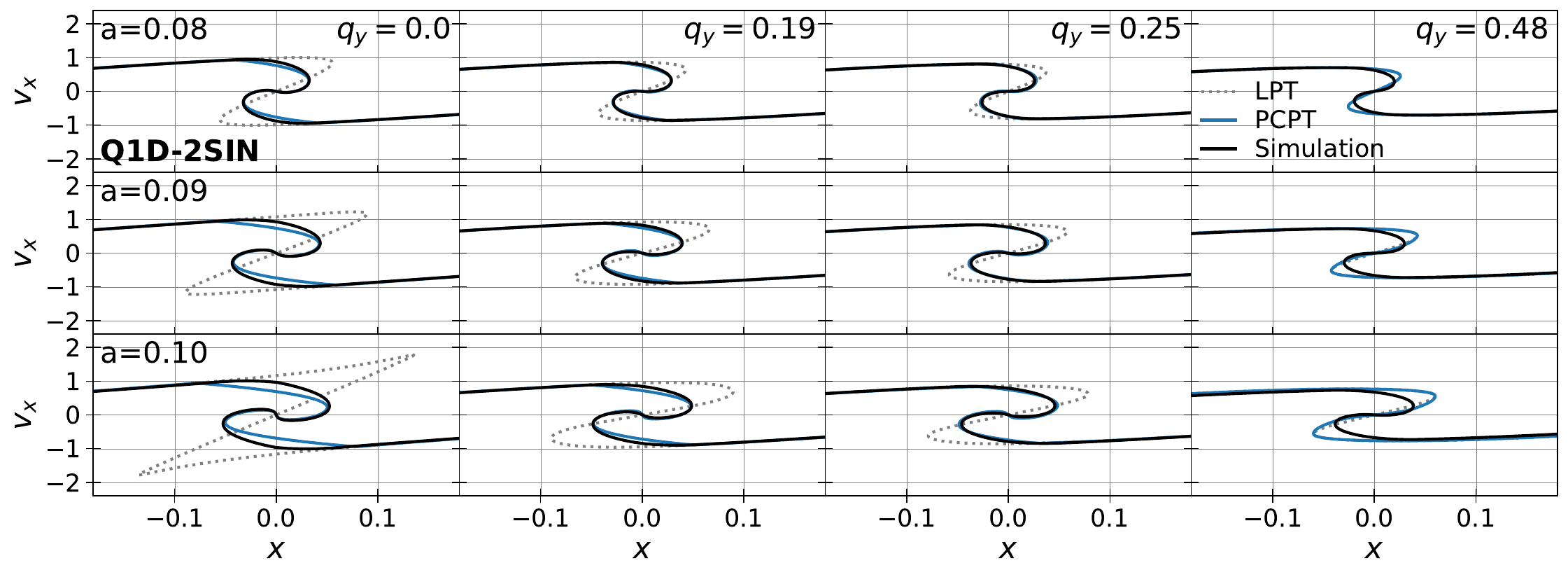}
\includegraphics[width=0.97\textwidth]{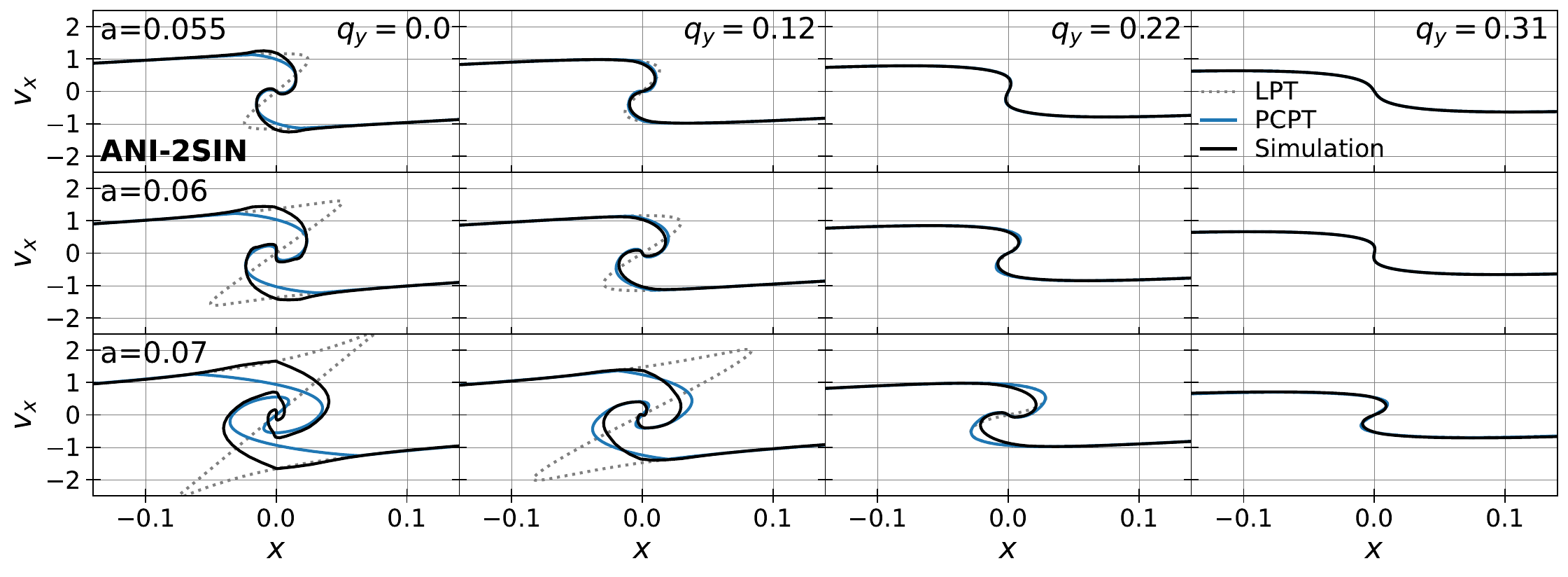}
\caption{Lagrangian phase-space slices $x(q_x,q_y)$-$v_{x}(q_x,q_y)$ for two-sine waves initial conditions. The top and bottom sets of panels correspond to Q1D-2SIN and ANI-2SIN cases, respectively. For each set of panels, time increases from top to bottom, and the fixed value of $q_y$ increases from left to right, that is, from the center to the outer part of the system.  The solid blue lines represent the analytical predictions based on PCPT with a 15LPT background flow (see Appendix~\ref{sec: app LPT dependence} for the effects of changing the LPT order $n$ on the solution), the black solid lines correspond to the measurements in the Vlasov-Poisson simulations, and the dotted lines give standard 15LPT predictions.}
\label{fig: x-vx 2SIN}
\end{figure*}
\begin{figure*}
\centering
\includegraphics[width=0.97\textwidth]{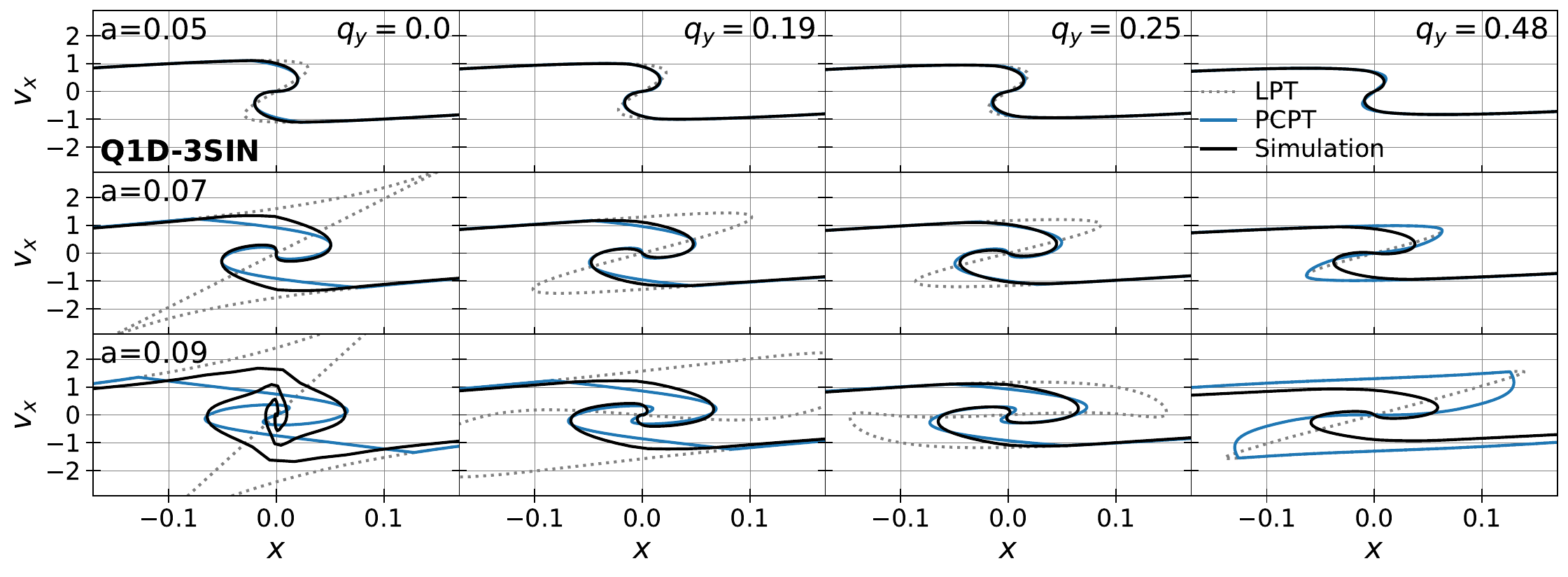}
\includegraphics[width=0.97\textwidth]{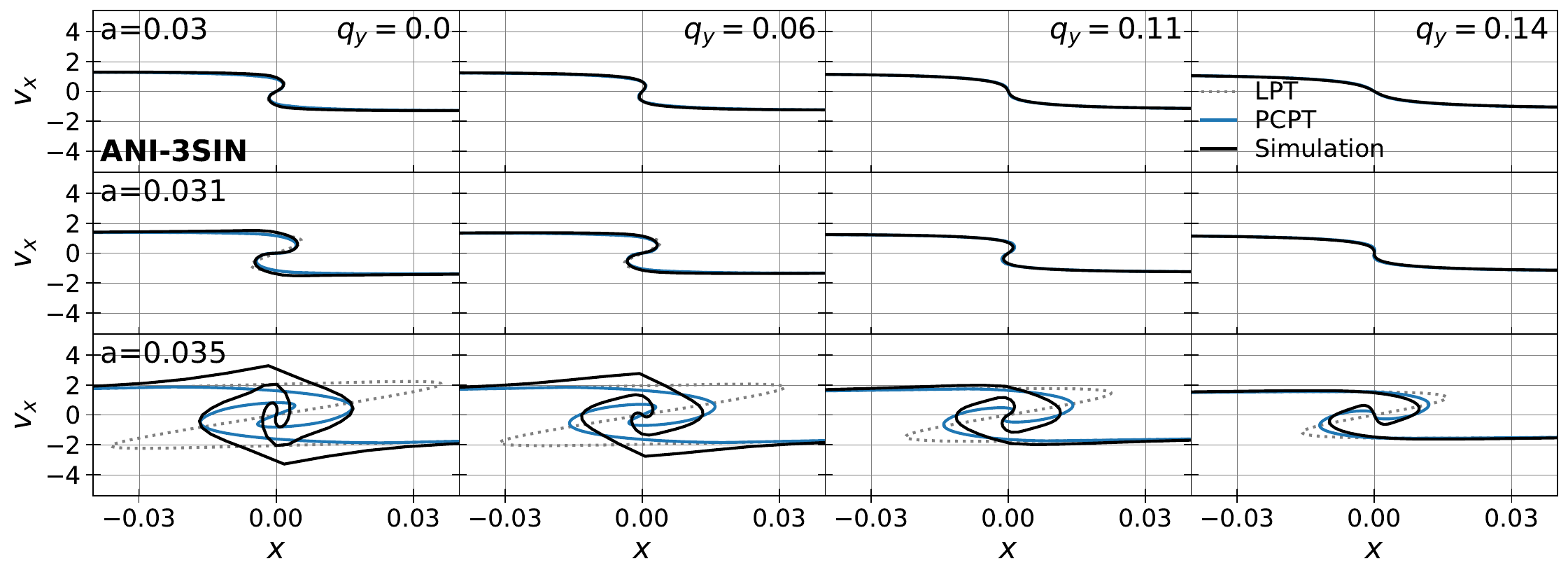}
\caption{Same as Fig.~\ref{fig: x-vx 2SIN} but for Q1D-3SIN (top panels) and ANI-3SIN cases (bottom panels), for $q_{z}=0$.}
\label{fig: x-vx 3SIN}
\end{figure*}

In Figs.~\ref{fig: x-vx 2SIN} and \ref{fig: x-vx 3SIN}, we examine Lagrangian phase-space slices, $x$-$v_{x}$, after the first shell crossing, by plotting functions $x(\tau,\bm{q})$ and $v_{x}(\tau,\bm{q})$ varying $q_{x}$ while fixing $q_{y}$ and $(q_{y},q_{z})$ for the two- and three-sine waves cases, respectively. We clearly see that the analytical predictions based on PCPT reproduce well the phase-space structure in the simulations for all the initial conditions even well after the first shell crossing. The post-collapse treatment is even able to approximate the second shell-crossing structure along $x$ direction, as already known in the one-dimensional case~\cite{2017MNRAS.470.4858T}. Strikingly, the snapshots at the times $a=0.09$ for Q1D-3SIN and $a=0.035$ for ANI-3SIN (bottom four panels in each figure) are beyond shell crossing along the $y$-axis (see Table~\ref{tab: ICs}), but PCPT is still able to qualitatively capture the phase-space structure in the vicinity of the origin.  The analytical predictions based on standard LPT are also shown in the figures, and they are clearly not able to perform as well because of the missing backreaction from the multi-stream flow. 

Focusing on the systems with quasi-one-dimensional initial conditions (Q1D-2SIN and Q1D-3SIN) for large values of $q_{y}$, i.e., in the outer regions of the pancakes (see the bottom-right panels of the Q1D results in Figs.~\ref{fig: x-vx 2SIN} and \ref{fig: x-vx 3SIN}.), we see that the analytical predictions based on both PCPT and standard LPT deviate significantly and similarly from the simulation results. This behavior is due to the fact that, at such large values of $q_y$, the Taylor expansion (\ref{eq: expand x}) and the ballistic approximation (\ref{eq: ballistic0}) become inaccurate and the force feedback is underestimated for PCPT, which is unable to bring proper correction on the background flow. In fact, if there is shell crossing for values of $q_y$ so close to the boundary of the periodic box, the shape of the multistream region is not anymore an ellipsoid in Lagrangian space, but becomes an infinite periodic ribbon.  In this case, the linear approximation (\ref{eq: pancake y}) obviously fails when $q_y$ approaches sufficiently the edge of the box (same for $q_z$ in 3D), which is equivalent to say, in terms of catastrophe theory, that the time considered is too advanced compared to shell crossing time, since keeping only leading order in $q_y$ is a consequence of leading order in time with the ballistic approximation (\ref{eq: ballistic0}) (see Ref.~\cite{2023A&A...678A.168S} for details). These effects are mitigated in the ANI cases because the output times considered tend to be less dynamically advanced than in the Q1D case, making Eqs.~(\ref{eq: pancake x}), (\ref{eq: pancake y}) and (\ref{eq: pancake z}) more accurate due to the smaller extension of the multistream region in Lagrangian space.

Another potential issue not addressed in Figs.~\ref{fig: x-vx 2SIN} and~\ref{fig: x-vx 3SIN} is LPT convergence. Computing high-order LPT solutions is increasingly costly with perturbative order $n$, this is why we stop at $n=15$ in the present work, which somewhat limits the accuracy and the predictive power of PCPT.  Additionally, even though it is known to behave smoothly for small enough times
(see, e.g., Refs.~\cite{2014JFM...749..404Z,2015MNRAS.452.1421R}), the LPT series has a finite radius of convergence and can diverge after shell crossing. How fast this happens depends on the nature of initial conditions (see e.g., Refs.~\cite{2021MNRAS.501L..71R,2023PhRvD.108j3513R} and references therein). The convergence of the LPT series also depends on position: it is expected to be worst around local peaks of the initial density field, but also around local minima by symmetry~\cite{2011MNRAS.410.1454N,2023PhRvD.107b3515R}. 
Note finally that it is difficult to tell whether the non-convergence is due to the LPT series does not truly converge, or simply because the LPT expansion order $n$ is too small, unless one carefully examines the behavior of the perturbative series in detail, which we do not do here.
Convergence with perturbative order $n$ is however briefly addressed in Appendix~\ref{sec: app LPT dependence}. One important conclusion of this appendix is that PCPT improves convergence in the multistream region, that is, the calculation of the force field with the backreaction in the multistream region, not only improves over LPT, but also converges better with order~$n$. This is of course true only as long as the PCPT prescription is valid (so no improvement is observed for large values of $|q_y|$ in the Q1D cases).

\subsubsection{Eulerian density slices}

\begin{figure*}
\centering
\includegraphics[width=0.49\textwidth]{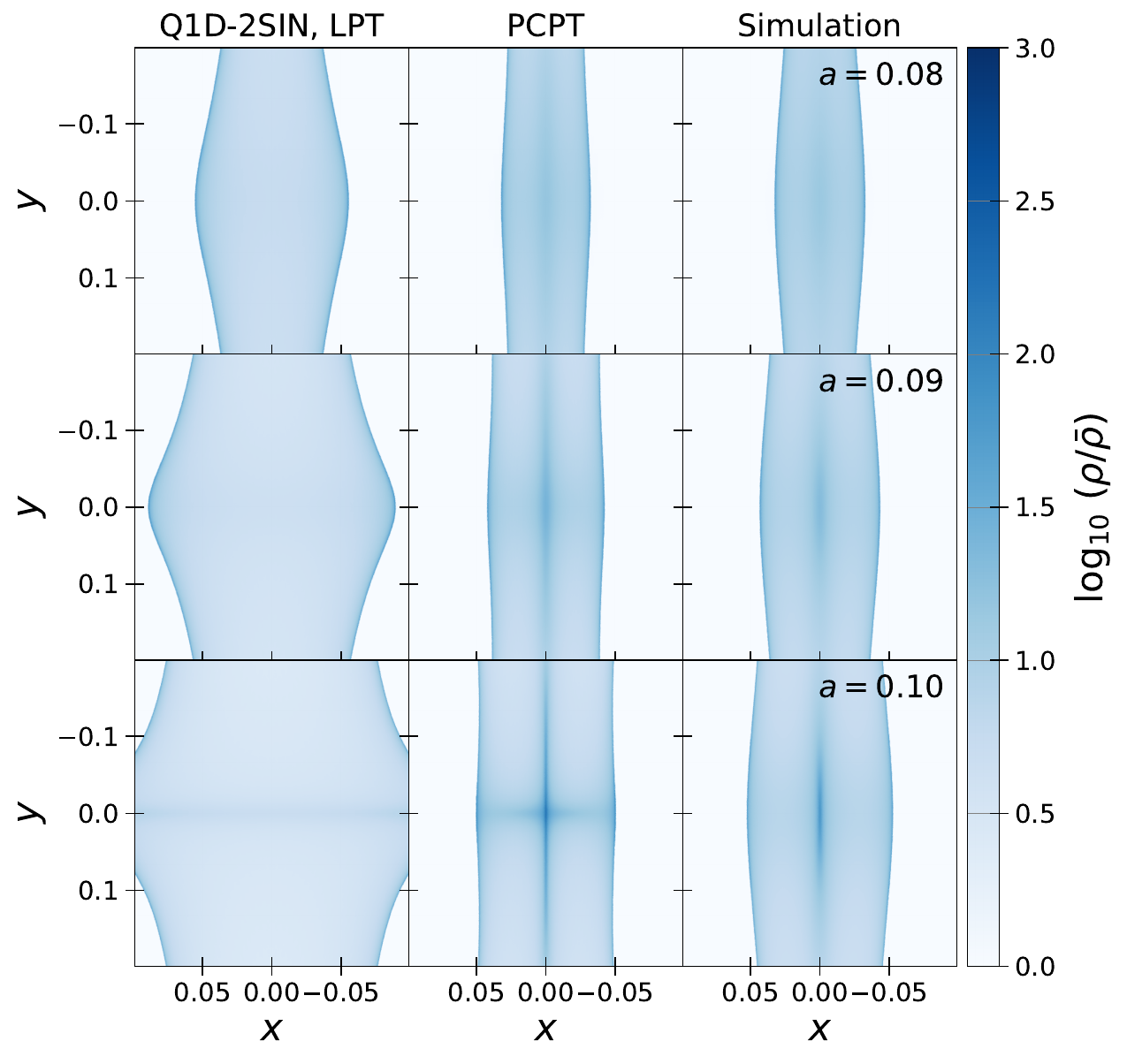}
\includegraphics[width=0.49\textwidth]{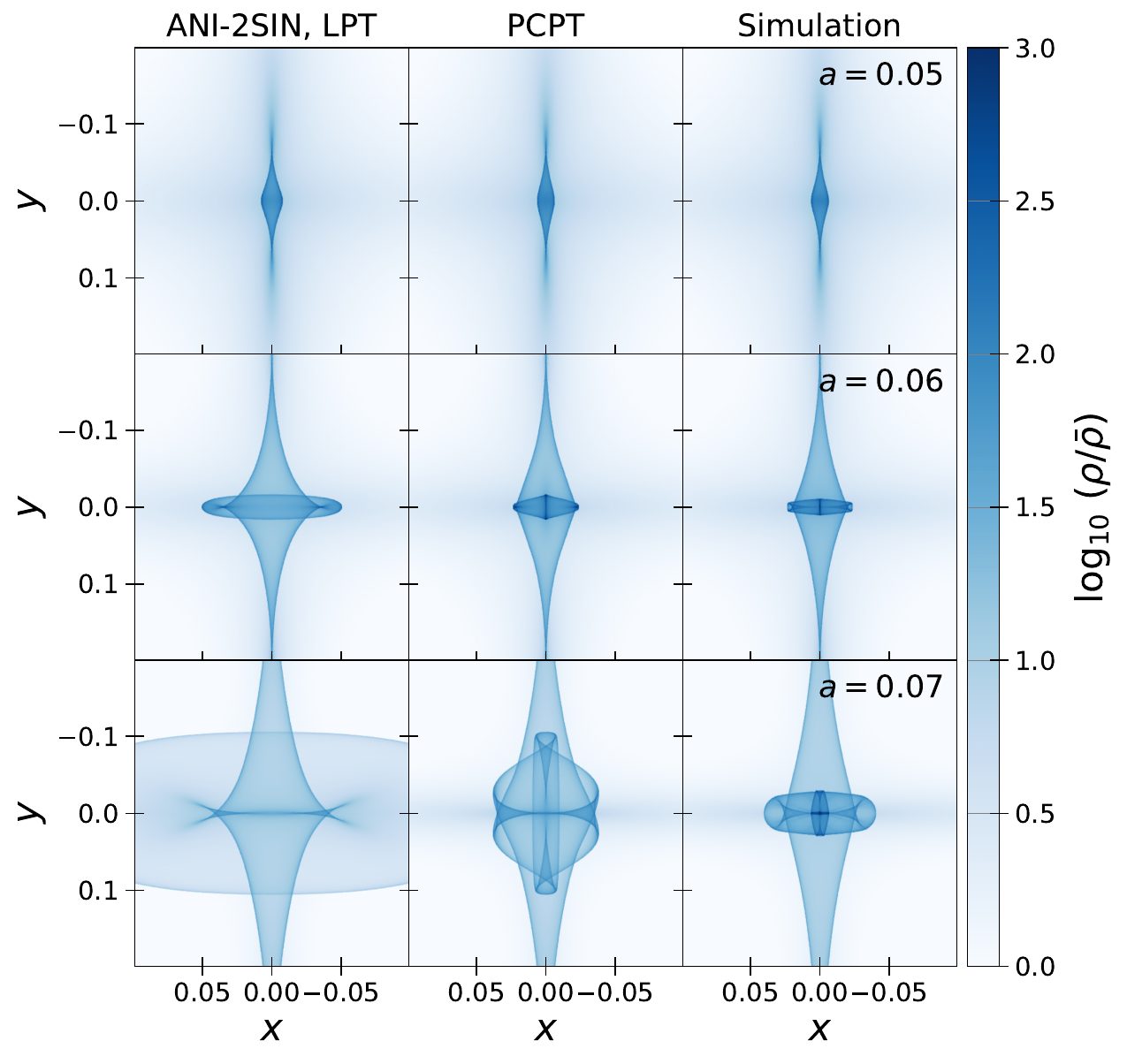}
\includegraphics[width=0.49\textwidth]{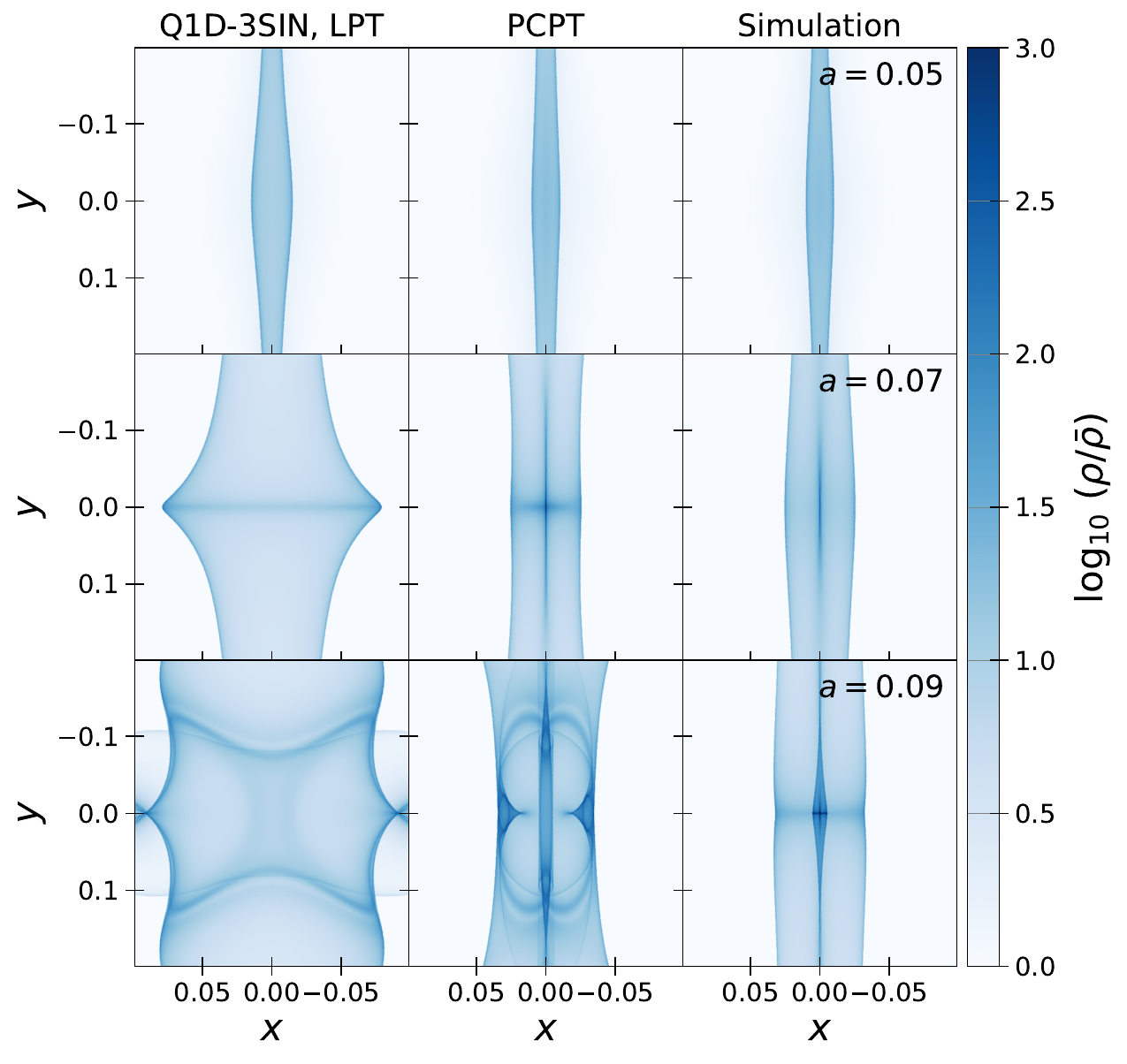}
\includegraphics[width=0.49\textwidth]{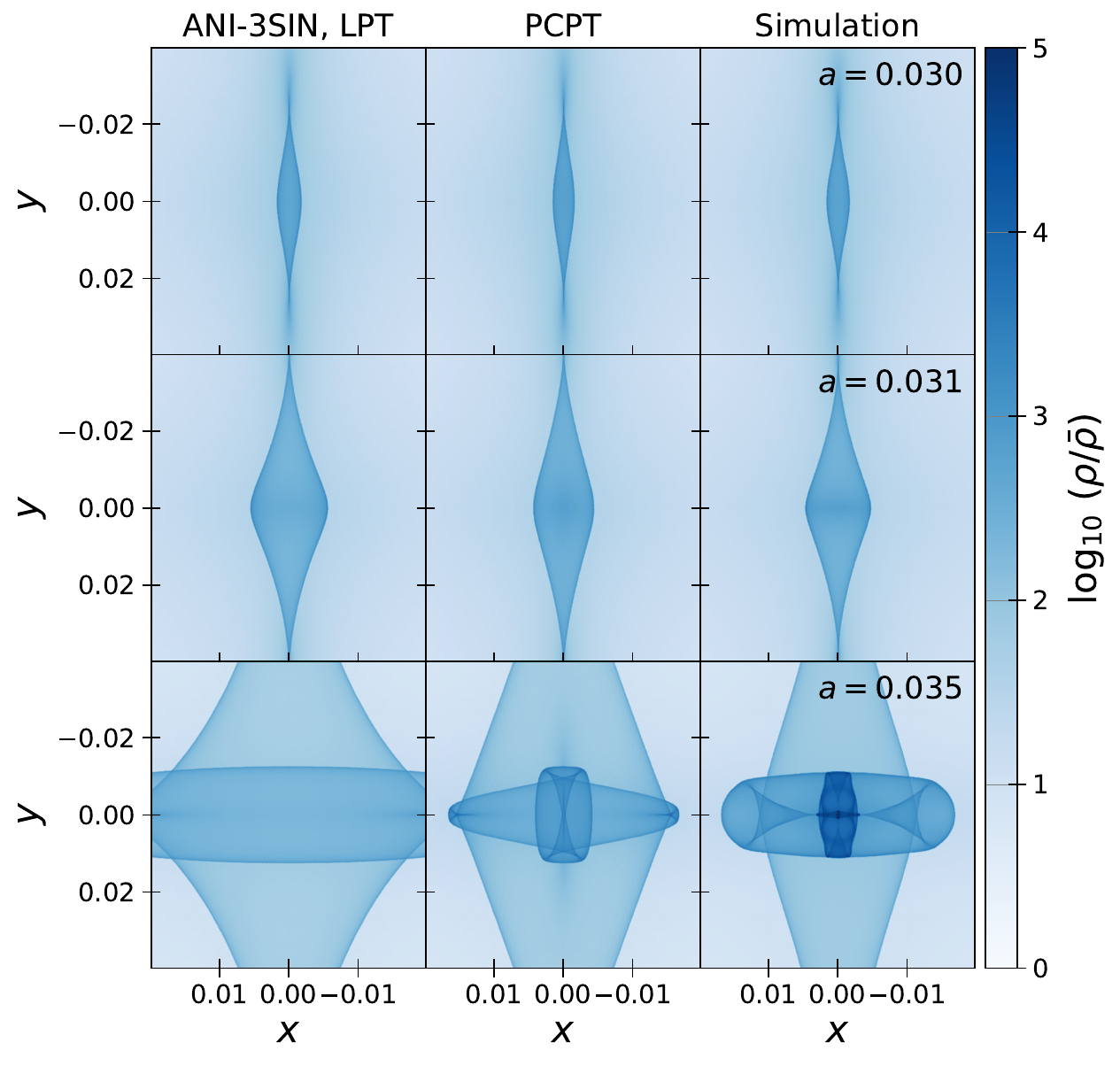}
\caption{Density slices at times indicated in the figures for two- and three-sine waves initial conditions. Top-left, bottom-left, top-right, and bottom-right panels, show density slices for Q1D-2SIN, Q1D-3SIN, ANI-2SIN, and ANI-3SIN cases, respectively. In each panel, from left to right, we present the results from standard 15LPT, PCPT with a 15LPT background flow, and the Vlasov-Poisson simulations, respectively. We note that for three-sine waves initial conditions, we show the two-dimensional slice with $z=0$.}
\label{fig: density}
\end{figure*}

Finally, to further understand how well PCPT is able to describe the multi-stream region and its caustic structure, Eulerian density slices are shown on Fig.~\ref{fig: density}. Their visual inspection confirms the analyses of the phase-space diagrams. Shortly after collapse (top lines of each group of panels), PCPT predictions clearly improve over standard LPT predictions. They reproduce accurately the simulation results, with a narrower extension of the pancake along $x$-axis thanks to an appropriate treatment of the backreaction of the force field. Later on the agreement remains good from the qualitative point of view, especially around the center of the system, even though the mismatch between theory and simulation becomes significant at late times (bottom lines of each group of panels). Interestingly, looking at the middle panels of the Q1D cases, standard LPT, and as a consequence, PCPT, predict earlier shell-crossing along the $y$-direction compared to the simulation measurements (see also Table~\ref{tab: ICs}). The reason for this counterintuitive result in the shell-crossing time along the $y$-direction remains to be understood. Clearly, because of the missing backreaction, non-linear couplings between axes of the motion are described incorrectly in the standard LPT case beyond first shell-crossing. While in Ref.~\cite{2023A&A...678A.168S}, it was found that the force field orthogonal to the direction of collapse was well described by LPT, this is only true shortly after shell-crossing. Furthermore, at later times, convergence of LPT can be questioned (see Appendix \ref{sec: app LPT dependence}), which obviously has some consequences on the estimate of shell-crossing time along $y$ axis. To resolve this problem, one has to study multistream dynamics beyond the approximations used in the present work with a better handling of LPT convergence, using e.g., UV completion, as discussed in next section. 

\section{Summary and discussion\label{sec: summary}}

In this paper, we have developed three-dimensional post-collapse perturbation theory (PCPT), that can describe the gravitational dynamics of cold dark matter (CDM) in the post-collapse regime, i.e., after the first shell crossing. Building on PCPT in one dimension~\cite{2015MNRAS.446.2902C,2017MNRAS.470.4858T}, we generalized the method to three dimensions, focusing on the pancake collapse of a CDM proto-halo. Pancake collapse represents the earliest and fundamental stage in the hierarchical formation of cosmic structures, where matter first undergoes gravitational collapse along a single direction to form thin, sheet-like over-densities. Since this configuration serves as a universal building block for more complex structures such as filaments and halos, understanding the post-collapse dynamics of pancakes is an essential step for describing the first stages of multistream dynamics of large-scale structures.

The approach underlying PCPT is based on (i) expanding an arbitrary pre-collapse flow in terms of the Lagrangian coordinates around the first shell-crossing position, (ii) deriving an expression of the gravitational force along the collapse direction with the one-dimensional Green function by exploiting the thinness of the pancake, and (iii) perturbatively computing the backreaction to correct the background flow along the collapse direction. The resulting expressions given in Sec.~\ref{sec: formulation} allow us to correct both the Eulerian position and velocity fields. A key technical advance is separating scales between the thinness of a pancake along its collapse axis and its much larger lateral extent. Exploiting this property allows us to asymptotically reduce the three-dimensional Poisson equation to an effectively one-dimensional form along the collapse direction while keeping transverse coordinates as parameters. As a result, we can utilize the PCPT prescription in one-dimensional cosmology~\cite{2015MNRAS.446.2902C,2017MNRAS.470.4858T} even in three-dimensional configurations.

We validated our theoretical framework by comparing its predictions with measurements in high-resolution Vlasov-Poisson simulations performed with the public code \texttt{ColDICE}~\cite{2016JCoPh.321..644S}, in Einstein-de Sitter universe (although our framework holds for $\Lambda$CDM universe), for two- and three-sine waves initial conditions, which are compatible with the pancake collapse condition. By varying the amplitude ratios of the sine waves, we investigated both quasi-one-dimensional (Q1D) and anisotropic cases (ANI). We set the 15th-order Lagrangian perturbation theory (LPT) solutions as the background flow in PCPT, and compared the theoretical predictions with the simulation results, as well as standard LPT predictions. These comparisons demonstrated that, overall, PCPT accurately reproduces the phase-space and density structures in the multi-stream regions and their vicinity, especially shortly after collapse inside and around the first shell-crossing locations, at variance with standard LPT which exhibits significant deviations from the simulations measurements.

Particularly interesting implications of our results lie in the potential of PCPT to bridge the early post-collapse regime with the emergent self-similar behavior of dark matter halos~\cite{1984ApJ...281....1F,1985ApJS...58...39B}. Recent 2D Vlasov-Poisson simulations~\cite{2025A&A...697A.218P} have demonstrated that, after only a few shell crossings, halos seeded by the same initial conditions as in our setups begin to trace self-similar trajectories. Our analytic framework may serve as a step towards connecting pre-collapse dynamics with the long-term dynamical attractors of collisionless self-gravitating systems. Exploring this connection further promises to yield deeper insight into the protohalo profiles~\cite{1991ApJ...382..377M,2005Natur.433..389D,2014ApJ...788...27I,2018MNRAS.473.4339O,2018PhRvD..97d1303D,2018PhRvD..98f3527D,2021A&A...647A..66C,2023MNRAS.518.3509D,2023JCAP...10..008D} and also the universal Navarro-Frenk-White (NFW) profile~\cite{1996ApJ...462..563N} measured in CDM simulations.

While the present formulation of PCPT for the three-dimensional pancake collapse already bridges the gap between pre-collapse regime and the early multi-stream regime, several important caveats remain:
\begin{itemize}
\item First, the current formulation focuses exclusively on pancake collapse and assumes a symmetric structure, which limits its applicability to more complex initial conditions, e.g., in the presence of tidal effects or mergers. 
\item Second, our perturbative correction is applied only along the primary collapse $x$-direction. Because the transverse $(y, z)$ motions are still taken from the uncorrected LPT background flow, the acceleration, which remains accurate only shortly after first shell-crossing, tends to be over-estimated at later times in transverse directions for Q1D cases and therefore PCPT (as well as LPT) predicts the shell-crossing along the $y$-axis earlier than the simulations.
\item Third, another limitation comes from the modelling of the background flow, i.e., high-order LPT in our analyses. After collapse, the convergence of LPT soon breaks down, even in underdense regions, potentially propagating unphysical features into the PCPT predictions. To address at least partly this issue, an improved treatment of the background flow via, for instance using UV completion approach~\cite{2023PhRvD.108j3513R,2023PhRvD.107b3515R}, together with a detailed understanding of the convergence radius of the LPT series, could enhance the robustness of PCPT.
\item Fourth, the accuracy of the Taylor expansion (\ref{eq: expand x}) and the ballistic approximation (\ref{eq: ballistic0}) we use to compute the PCPT solution deteriorate significantly after shell crossing, particularly so in the outskirts of the evolving pancake for the Q1D case. Although this is naturally expected due to the perturbative nature of our calculation, one can improve on these issues, by simultaneously increasing the orders of the expansion in $\bm{q}$ and in $\tau-\tau_c$. This is a complex procedure which requires an iterative process. Indeed, higher order in $\tau-\tau_{\rm c}$ first requires prior calculation of the leading order background correction on the force field.\footnote{Sect.~\ref{sec: formulation} proposes expressions of the force field at next to leading (linear) order in $\tau-\tau_{\rm c}$, but the coefficients $\mathcal{A}^{(1)}(\bm{q})$ and $\mathcal{B}^{(1)}(\bm{q})$ for the $\tau-\tau_{\rm c}$ contribution are at best approximate, even at the asymptotic level. As a result we ignored such corrections in the final expressions for the position and the velocity given in Eqs.~(\ref{eq: Delta x out})--(\ref{eq: Delta u in}).} Such higher order correction is needed to justify the introduction of higher order terms in the Taylor expansion in $\bm{q}$. Another possible way to improve on PCPT is to perform partial Taylor expansion in $\bm{q}$ space orthogonally to the $A_3^{+}$ ridges as defined in catastrophe theory \citep[see][and references therein]{2018JCAP...05..027F}, instead of restricting to expansion around the first shell crossing point. 
\end{itemize}
All these potential improvements are left for future works. These include the extension of the PCPT to more general initial conditions, such as those given by a random Gaussian field. This case is more relevant to the real universe.

\begin{acknowledgments}
This work was supported in part by the Japan Society for the Promotion of Science (JSPS) Overseas Research Fellowships, the JSPS KAKENHI Grant Numbers JP23K19050 and JP24K17043 (SS), JP20H05861, JP23K20844 and JP23K25868 (AT), the French Doctoral school ED127, Astronomy and Astrophysics Ile de France (AP), the Programme National Cosmology et Galaxies (PNCG) of CNRS/INSU with INP and IN2P3, co-funded by CEA and CNES, the « action th{\'e}matique » Cosmology-Galaxies (ATCG) of the CNRS/INSU PN Astro, and by the “PHC Sakura” program (project number: 51203TL, grant number: JPJSBP120243208), implemented by the French Ministry for Europe and Foreign Affairs, the French Ministry of Higher Education and Research and the Japan Society for Promotion of Science (JSPS).  Numerical computation with \texttt{ColDICE} was carried out using the HPC resources of CINES (Occigen supercomputer) under the GENCI allocations c2016047568, 2017-A0040407568 and 2018-A0040407568. Post-treatment of ColDICE data was performed on HORIZON (subsequently INFINITY) cluster of Institut d'Astrophysique de Paris. We thank St\'ephane Rouberol for his advice and help while using this cluster. 
\end{acknowledgments}

\appendix

\section{Derivations of the expressions of force\label{sec: app Fx}}

Starting from the one-dimensional Poisson equation Eq.~(\ref{eq: 1D Poisson eq}), we derive the analytical expression of the gravitational force along the collapse direction ($x$-direction in the present case) for the pancake. The $x$-component of the force at $x=x(\tau,\bm{q})$ is given by solving equation (\ref{eq: 1D Poisson eq}) with the Green function method:
\begin{align}
F_{x}(\tau,\bm{x}) 
&= - \frac{3}{2}\Omega_{\rm m0}H^{2}_{0}
\frac{a}{B_{y}(\tau)B_{z}(\tau)}\notag \\
& \times 
\int{\rm d}x'\, \left| \frac{\partial x}{\partial q_{x}} \right|^{-1} \left( \partial_{x}G_{\rm 1D}(x,x')\right)
, \\
& \equiv 
- \frac{3}{2}\Omega_{\rm m0}H^{2}_{0}
\frac{a}{B_{y}(\tau)B_{z}(\tau)}
\mathcal{F}(\tau, \bm{q}) , \label{eq: Fx def}
\end{align}
where we define the Green function $G_{\rm 1D}(x,x')$ as \citep[see e.g.][]{2015MNRAS.446.2902C,2017MNRAS.470.4858T}:
\begin{align}
\partial_{x}G_{\rm 1D}(x,x') & = 
\frac{1}{2}\Bigl[ \Theta(x-x')
- \Theta(x'-x) \Bigr]
- \frac{1}{L}(x-x') .
\end{align}
The function $\mathcal{F}(\tau, \bm{q})$ is given by (see also Fig.~\ref{fig: schematic}):
\begin{widetext}
\begin{align}
\mathcal{F}(\tau, \bm{q})
& = Q_{-} - Q_{*} + Q_{+} - x(\tau,Q_{*}) , \notag \\
& =
\begin{cases}
-2q_{x} + \beta(\tau,q_{y},q_{z}) q_{x} - C_{x}(\tau) q_{x}^{3} 
&
\mbox{for $|q_{x}|  \leq Q_{{\rm c}}(\tau,q_{y},q_{z})$} , \\
q_{x} + \beta(\tau,q_{y},q_{z}) q_{x} - C_{x}(\tau) q_{x}^{3}
- {\rm sgn}(q_{x})\sqrt{3}\sqrt{ \hat{Q}_{{\rm c}}^{2}(\tau,q_{y},q_{z})-q_{x}^{2} }
& \mbox{for $Q_{{\rm c}}(\tau,q_{y},q_{z})<|q_{x}|\leq \hat{Q}_{{\rm c}}(\tau,q_{y},q_{z})$} , \label{eq: calF def}
\end{cases}
\end{align}
\end{widetext}
with
\begin{align}
\beta(\tau, q_{y},q_{z}) &= B_{x}(\tau) - \left( C_{y}(\tau) q_{y}^{2} + C_{z}(\tau) q_{z}^{2} \right) .
\end{align}
After substituting Eq.~(\ref{eq: calF def}) into Eq.~(\ref{eq: Fx def}), we expand the force expression $F_{x}(\tau,\bm{q})$ up to linear order in $\tau-\tau_{\rm sc}$, keeping the factor $\hat{Q}_{{\rm c}}(\tau, q_{y}, q_{z})$ in the square root term, to obtain the following expression
\begin{align}
F_{x}(\tau, \bm{q})
& \simeq -\frac{3}{2}\Omega_{\rm m0}H^{2}_{0}\frac{a(\tau_{\rm sc})}{B^{(0)}_{y} B^{(0)}_{z}}\tilde{\mathcal{F}}(\tau,\bm{q}) ,\label{eq: app Fx expand}
\end{align}
with
\begin{widetext}
\begin{align}
\tilde{\mathcal{F}}(\tau,\bm{q}) &=
\begin{cases}
\mathcal{A}^{(0)}(\bm{q}) + \mathcal{A}^{(1)}(\bm{q})  (\tau-\tau_{\rm sc})
&
\mbox{for $|q_{x}| \leq Q_{{\rm c}}(\tau,q_{y},q_{z})$} , \\
\mathcal{B}^{(0)}(\bm{q}) 
+ \mathcal{B}^{(1)}(\bm{q})  (\tau-\tau_{\rm sc})
- {\rm sgn}(q_{x})\sqrt{3}\sqrt{ \hat{Q}_{{\rm c}}^{2}(\tau,q_{y},q_{z})-q_{x}^{2} }
&
\mbox{for $Q_{{\rm c}}(\tau,q_{y},q_{z})<|q_{x}| \leq \hat{Q}_{{\rm c}}(\tau,q_{y},q_{z})$} .
\end{cases}
\end{align}
\end{widetext}
In the above, we further defined
\begin{align}
\mathcal{A}^{(0)}(\bm{q}) &= q_{x}(-2 + \beta^{(0)}(q_{y},q_{z}) - C^{(0)}_{x}q_{x}^{2}) , \label{eq: calF in 0}\\
\mathcal{A}^{(1)}(\bm{q}) &= q_{x}\Bigl[-2\alpha + \alpha \beta^{(0)}(q_{y},q_{z}) 
\notag \\
& + \beta^{(1)}(q_{y},q_{z}) - (C^{(1)}_{x}+C^{(0)}_{x}\alpha)q_{x}^{2})\Bigr] , \\
\mathcal{B}^{(0)}(\bm{q}) &= q_{x}(1 + \beta^{(0)}(q_{y},q_{z})
- C^{(0)}_{x}q_{x}^{2}) , \label{eq: calF out 0} \\
\mathcal{B}^{(1)}(\bm{q}) &= q_{x}\Bigl[ \alpha + \alpha \beta^{(0)}(q_{y},q_{z}) + \beta^{(1)}(q_{y},q_{z}) \notag \\
& - (C^{(1)}_{x}+C^{(0)}_{x}\alpha)q_{x}^{2}) \Bigr] ,  \label{eq: calF out 1}
\end{align}
with the coefficient $\alpha$ and the functions $\beta^{(0)}(q_{y},q_{z})$ and $\beta^{(1)}(q_{y},q_{z})$ given by
\begin{align}
\alpha & = a^{2}(\tau_{\rm sc})H(\tau_{\rm sc}) + \frac{B^{(1)}_{y}B^{(0)}_{z}+B^{(0)}_{y}B^{(1)}_{z}}{B^{(0)}_{y}B^{(0)}_{z}} ,\\
\beta^{(0)}(q_{y},q_{z}) &= - C^{(0)}_{y} q_{y}^{2} - C^{(0)}_{z} q_{z}^{2} ,\\
\beta^{(1)}(q_{y},q_{z}) &= B^{(1)}_{x} - C^{(1)}_{y} q_{y}^{2} - C^{(1)}_{z} q_{z}^{2} .
\end{align}
We can include the background force given in Eq.~(\ref{eq: bg force}), which gives negligible contributions, though, by adding the corresponding term:
\begin{align}
\mathcal{F}_\BG(\bm{q}) &= \frac{B^{(0)}_{y}B^{(0)}_{z}}{d}q_{x}
\Biggl[
\beta^{(0)}(q_{y},q_{z}) - C^{(0)}_{x}q_{x}^{2}
\notag \\
&
+ (\tau - \tau_{\rm sc}) \Bigl( \beta^{(1)}(q_{y},q_{z}) - C^{(1)}_{x}q_{x}^{2}
\notag \\
&
+ (-C^{(0)}_{x}q_{x}^{2}+\beta^{(0)})a^{2}(\tau_{\rm sc})H(\tau_{\rm sc}) \Bigr) 
\Biggr].
\end{align}
As discussed in the main text, in the above expressions for the force field, the $\tau-\tau_{\rm c}$ terms are only approximate and will be ignored in the final calculation of the PCPT position and velocity, exactly as in  Ref.~\cite{2017MNRAS.470.4858T} for the 1D problem, for which $\mathcal{F}_\BG(\bm{q})$ is null because $B_y^{(0)}=B_z^{(0)}=0$ in this latter case.

\section{Derivations of corrections to the background flow \label{sec: app derivations}}

Based on the force expression derived in Appendix~\ref{sec: app Fx}, we provide here additional technical details needed to compute the corrections to the background flow. In doing so, we need to perform the time-integrals in Eqs.~(\ref{eq: def Delta u out}), (\ref{eq: def Delta x out}), (\ref{eq: def Delta u in 2}),  and (\ref{eq: def Delta x in 2}), using the leading order in $\tau-\tau_{\rm c}$ of Eq.~(\ref{eq: app Fx expand}) for the expression of the force, i.e., dropping terms proportional to $\tau-\tau_{\rm c}$ by setting $\mathcal{A}^{(1)}(\bm{q})=\mathcal{B}^{(1)}(\bm{q})=0$.

Integral (\ref{eq: def Delta u out}) can be easily calculated to obtain Eq.~(\ref{eq: app Delta u out}) by exploiting the analytical expression of the following $\tau'$-integral,
\begin{align}
&
\int^{\tau}_{\hat{\tau}_{{\rm c}}(\bm{q})}{\rm d}\tau'\,
\left( \hat{Q}^{2}_{{\rm c}}(\tau',q_{y},q_{z})-q_{x}^{2} \right)^{1/2}
\notag \\
&
\qquad =
\frac{1}{12}\kappa^{(1)}  \left( \hat{Q}_{{\rm c}}^{2}(\tau,q_{y},q_{z})-q_{x}^{2}\right)^{3/2}.
\end{align}
Additional integration other time and using the analytical expression of the following $\tau'$-integral:
\begin{align}
&
\int_{\hat{\tau}_{{\rm c}}(\bm{q})}^{\tau}{\rm d}\tau'\,
\left( \hat{Q}_{{\rm c}}^{2}(\tau',q_{y},q_{z})-q_{x}^{2} \right)^{3/2}
\notag \\
& \qquad 
= \frac{\kappa^{(1)} }{20}\left( \hat{Q}_{{\rm c}}^{2}(\tau,q_{y},q_{z})-q_{x}^{2} \right)^{5/2} ,
\end{align}
we can compute integral (\ref{eq: def Delta x out}) to obtain Eq.~(\ref{eq: Delta x out}).

Integral (\ref{eq: def Delta u in 2}) can be easily computed, to obtain Eq.~(\ref{eq: Delta u in}), by using analytic expressions of the following $\tau'$-integrals:
\begin{align}
&
\int^{\tau_{{\rm c}}(\bm{q})}_{\hat{\tau}_{{\rm c}}(\bm{q})}{\rm d}\tau'\, 
= \frac{3}{8}\kappa^{(1)} (q_{y},q_{z})q_{x}^{2}
, \\
&
\int^{\tau_{{\rm c}}(\bm{q})}_{\hat{\tau}_{{\rm c}}(\bm{q})}{\rm d}\tau'\, 
\left( \hat{Q}_{{\rm c}}^{2}(\tau,q_{y},q_{z})-q_{x}^{2} \right)^{1/2}
= 
\frac{\sqrt{3}}{4}\kappa^{(1)}  |q_{x}|^{3} . \label{eq:thisone}
\end{align}
Finally, we exploit the analytical expressions of the following $\tau'$-integrals:
\begin{align}
&
\int^{\tau_{{\rm c}}(\bm{q})}_{\hat{\tau}_{{\rm c}}(\bm{q})}{\rm d}\tau'\, \left( \tau - \hat{\tau}_{{\rm c}}(\bm{q}) \right)
= \frac{9}{128}\left(\kappa^{(1)}(q_{y},q_{z})\right)^{2}q_{x}^{4} 
, \\
&
\int^{\tau_{{\rm c}}(\bm{q})}_{\hat{\tau}_{{\rm c}}(\bm{q})}{\rm d}\tau'\,
\left( \hat{Q}_{{\rm c}}^{2}(\tau,q_{y},q_{z})-q_{x}^{2}\right)^{3/2}
\notag \\
& \qquad
= \frac{9}{20}\sqrt{3}\kappa^{(1)} (q_{y},q_{z})|q_{x}|^{5} ,
\end{align}
to perform additional time integration in Eq.~(\ref{eq: def Delta x in 2}) and we obtain Eq.~(\ref{eq: Delta x in}).

\section{Dependence of PCPT results on the LPT order of the background flow\label{sec: app LPT dependence}}

\begin{figure*}
\centering
\includegraphics[width=0.97\textwidth]{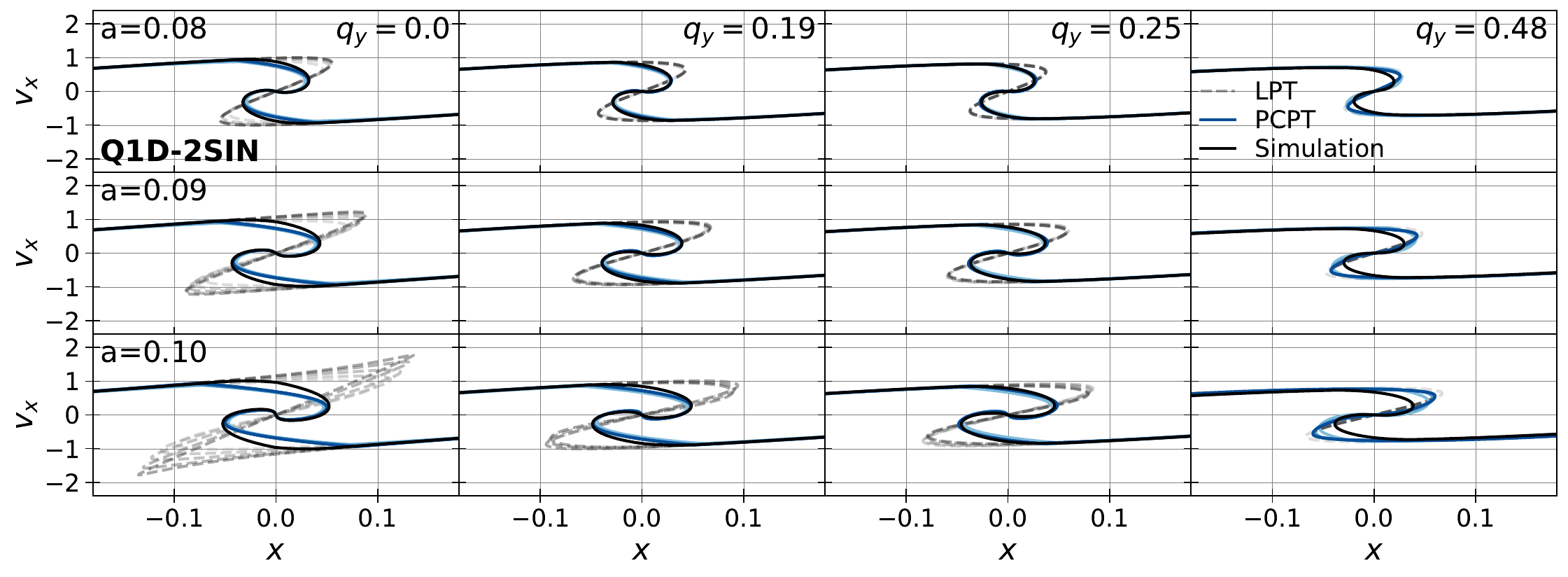}
\includegraphics[width=0.97\textwidth]{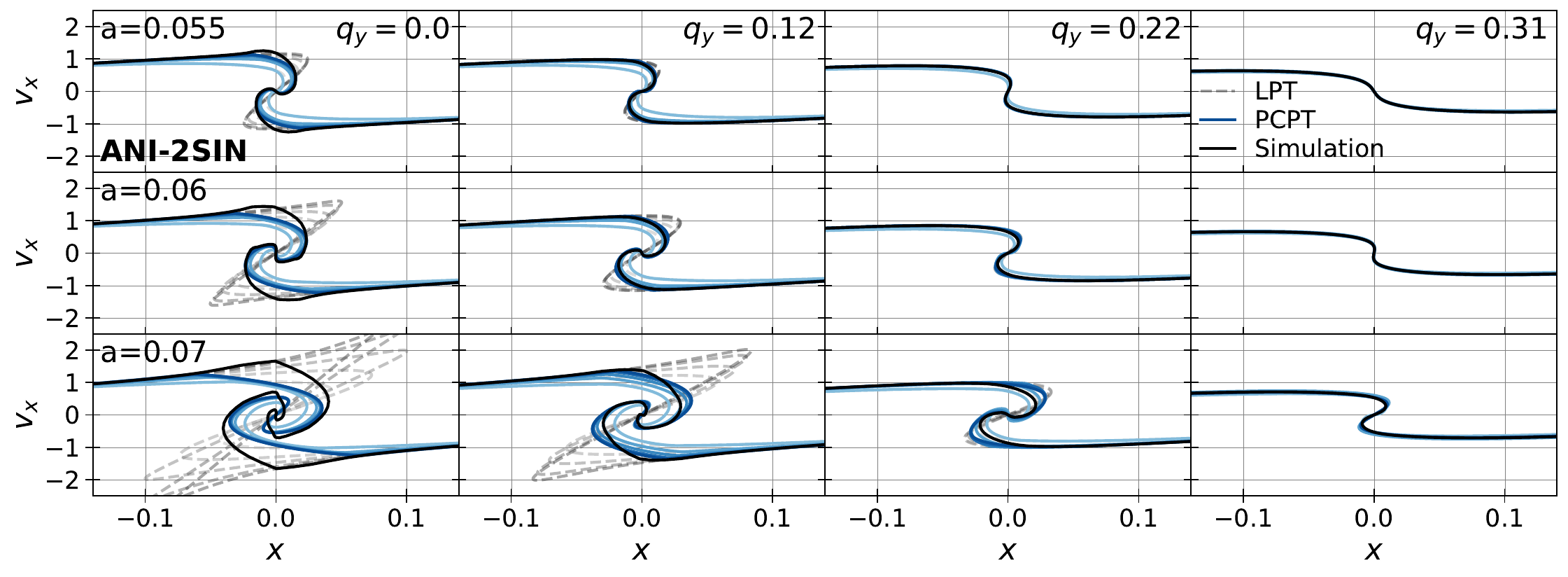}
\caption{Lagrangian phase-space slices $x(q_x,q_y)$-$v_{x}(q_x,q_y)$ in the 2D configurations, similarly as in Fig.~\ref{fig: x-vx 2SIN}. Here, we test how the results depend on the LPT order $n$. From light to dark blue, PCPT is shown with the background flow modelled with 2LPT, 5LPT, 9LPT, 13LPT, and 15LPT, respectively. The standard LPT results are also displayed in grey dashed, while the black curve corresponds to the Vlasov simulations.}
\label{fig: x-vx nLPT 2SIN}
\end{figure*}

\begin{figure*}
\centering
\includegraphics[width=0.97\textwidth]{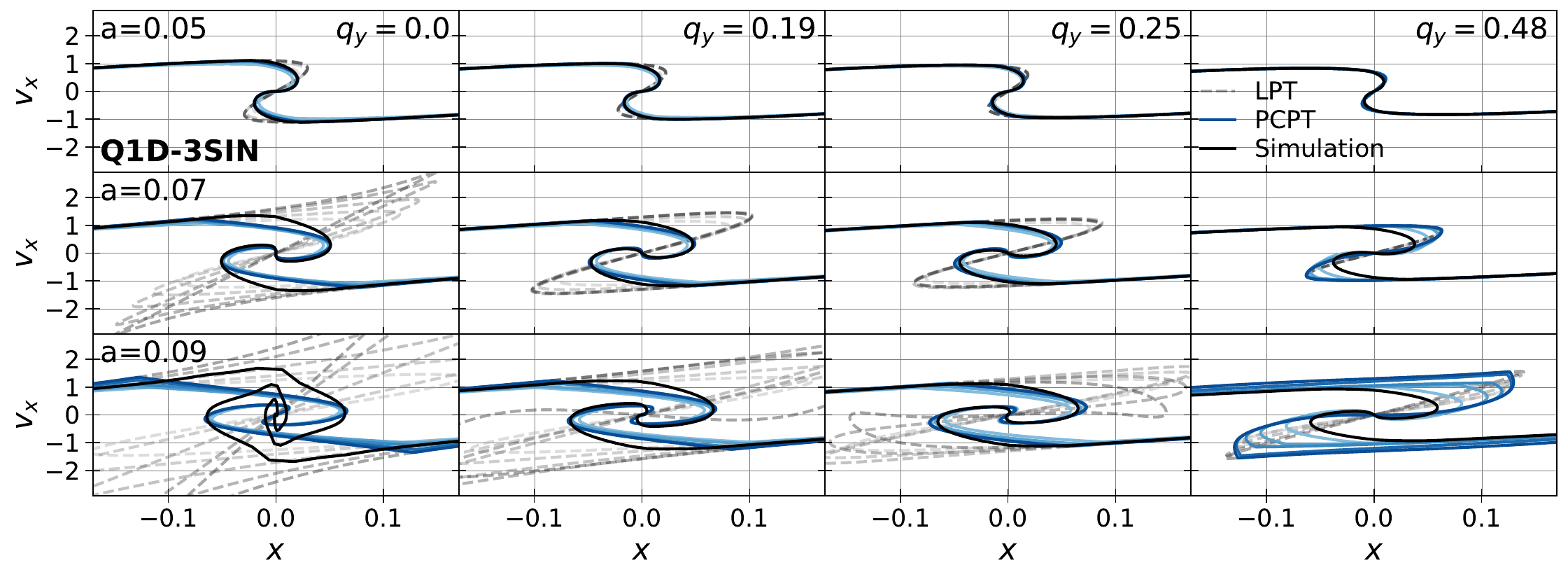}
\includegraphics[width=0.97\textwidth]{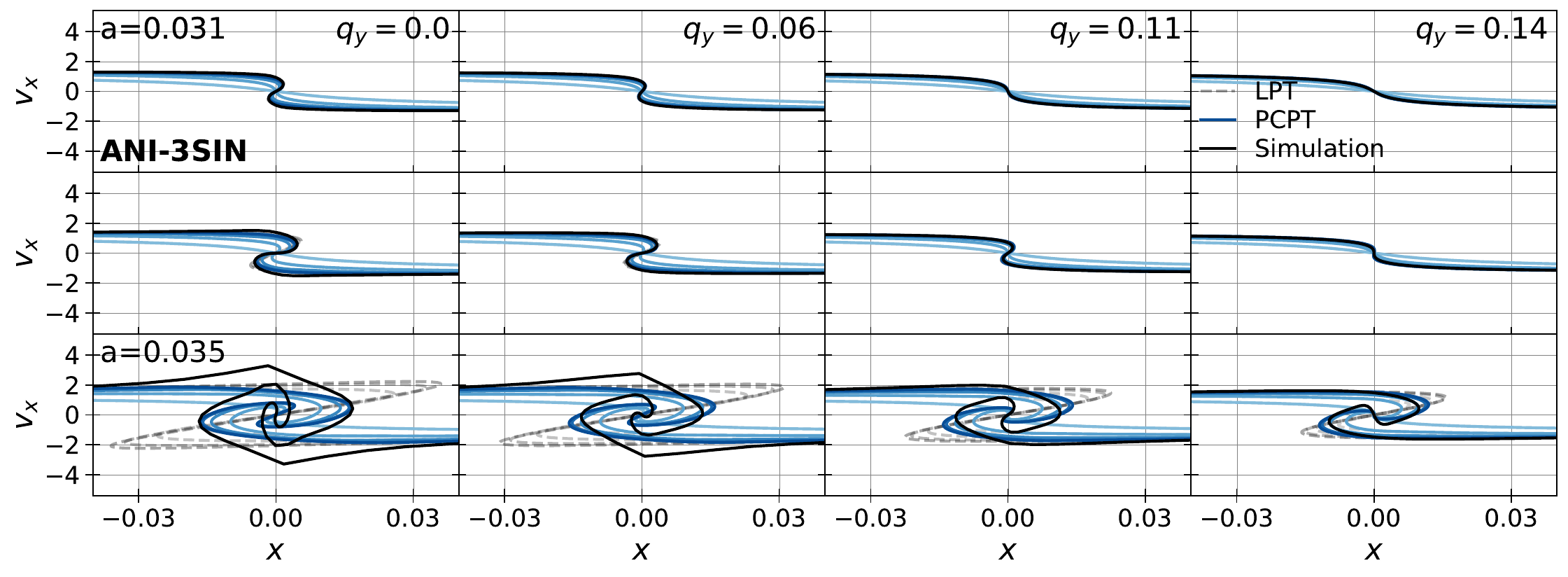}
\caption{Same as Fig.~\ref{fig: x-vx nLPT 2SIN}, but for the 3D cases.}
\label{fig: x-vx nLPT 3SIN}
\end{figure*}

In this Appendix, we test the performances of PCPT with respect to the LPT order used to model the background flow. More specifically, in addition to 15LPT, we consider $n$LPT with $n=2,$ 5, 9 and 13, as illustrated by the phase-space diagrams displayed in Figs.~\ref{fig: x-vx nLPT 2SIN} and \ref{fig: x-vx nLPT 3SIN} for the 2SIN and 3SIN cases, respectively.

Before commenting on convergence with $n$, we must be aware that PCPT behaves poorly for the largest value of $q_y$ considered in the Q1D cases, as illustrated by right parts of the upper group of panels in figures \ref{fig: x-vx nLPT 2SIN} and \ref{fig: x-vx nLPT 3SIN}, while it performs much better when approaching the center of the system or for ANI initial conditions.  This is not related to LPT convergence and is merely a limit of the Taylor expansion (\ref{eq: expand x}) and the linear approximations (\ref{eq: pancake y}) and (\ref{eq: pancake z}), which fail for such a large value of $q_y$ (or $q_z$) nearly on the edge of the periodic box, as discussed in detail in Sect.~\ref{sec: comparison}. This issue put aside, we can see that convergence is clearly not achieved for $n=15$ in the bottom part of each group of panels in the figures, particularly the bottom left part.  Interestingly, when they are applicable, PCPT corrections improve greatly the convergence of the final solution with perturbative order, compared to pure LPT, as clearly visible for instance on lower left part of each group of panels in the figures. This important property of PCPT might turn to be extremely useful when computing statistical properties of the matter distribution.

\bibliography{ref}
\end{document}